\documentclass[12pt]{article}
\usepackage{amssymb,amsmath,amsfonts}
\usepackage{bm}
\usepackage{cite}
\usepackage{mathrsfs}
\usepackage{slashed}
\usepackage{graphicx}
\usepackage{hyperref}
\usepackage{verbatim}
\usepackage{color}
\usepackage{enumerate}
\usepackage{authblk}
\usepackage[hang,flushmargin]{footmisc}
\usepackage{lipsum}
\usepackage[font=footnotesize]{caption}
\usepackage{soul}
\usepackage[utf8]{} 
\usepackage{empheq}
\usepackage{bigints}

\usepackage[makeroom]{cancel}
\usepackage[normalem]{ulem}

\usepackage{atbegshi,picture}
\usepackage{lipsum}
\usepackage[T1]{fontenc}

\title{\bf Holography of dislocations and ring defects in Einstein-Gauss-Bonnet AdS gravity }

\author{Vladimir  Juri\v ci\'c$^{1,}$\thanks{{vladimir.juricic@usm.cl}}\;,  Olivera Miskovic$^{2,}$\thanks{olivera.miskovic@pucv.cl}\;, 
Francisca Ram\'irez Carrasco$^{1,2,}$\thanks{francisca.ramirez@alumnos.usm.cl}
\bigskip\\
{\small $^1$  Departamento de F\'isica, Universidad T\'ecnica Federico Santa Mar\'ia, Casilla 110, Valpara\'iso, Chile}\smallskip\smallskip\\
{\small $^2$ Instituto de F\'\i sica, Pontificia Universidad Cat\'olica de Valpara\'\i so,}\\ {\small Avda.~Universidad 330, Curauma, Valpara\'{\i}so, Chile.}
}

\newcommand{\de}{\mathrm{d}}

\definecolor{blue-violet}{rgb}{0.54, 0.17, 0.89}
\definecolor{PineGreen}{cmyk}{0.92, 0, 0.59, 0.25}
\definecolor{OliveGreen}{cmyk}{0.64, 0, 0.95, 0.40}
\definecolor{RawSienna}{cmyk}{0, 0.72, 1, 0.45}
\definecolor{Gray}{cmyk}{0, 0, 0, 0.50}
\definecolor{MidnightBlue}{cmyk}{0.98, 0.13, 0, 0.43}
\definecolor{Orange}{cmyk}{0, 0.61, 0.87, 0}
\definecolor{LimeGreen}{cmyk}{0.50, 0, 1, 0}
\definecolor{Green}{cmyk}{1, 0, 1, 0}

\numberwithin{equation}{section}

\begin{document}

\maketitle

\begin{abstract} We study torsional topological defects in Einstein-Gauss-Bonnet gravity in ($4+1$)-dimensional anti-de Sitter spacetime. In the holographic interpretation, these correspond to crystalline dislocation defects associated with the discrete lattice translational symmetry. The Gauss-Bonnet coupling is fixed at the Chern-Simons point. By solving the equations of motion through an asymptotic expansion near the boundary, we show that the dual ($3+1$)-dimensional theory admits axially symmetric solutions. These solutions describe holographic materials with dislocation defects at finite temperature, encoded by a black hole in the bulk. At the same time, they feature ring-shaped defects arising from the background Riemann-Cartan geometry, characterized by nontrivial Burgers vectors. We also discuss the possible appearance of an odd-parity Abelian holographic anomaly, proportional to the Nieh-Yan invariant. Our results motivate further studies of holographic defects using bulk gravitational theories and support the view that torsion provides a holographic counterpart of crystalline dislocation defects.

\end{abstract}

\tableofcontents

\section{Introduction}

In gravitational theories, the search for topological defect solutions is natural since topology and geometry themselves become dynamical. From objects such as magnetic monopoles, cosmic strings, and domain walls
\cite{Kibble:1976sj,Vilenkin:1984ib,Hanany:2005bq} to various solitonic solutions in gravity, e.g., the Taub-NUT spacetime \cite{Taub:1950ez,Newman:1963yy} and the Eguchi-Hanson instanton \cite{Eguchi:1978xp}, these solutions have attracted attention due to their stability and their role as generalizations of familiar gravitational objects. They find applications in early universe cosmology \cite{Kibble:1976sj,Vilenkin:1984ib}, in quantum gravity \cite{Ashtekar:1997yu,Maldacena:2001xj}, topological phases \cite{volovik2003universe,Teo-Hughes-2017}, and in holography \cite{Li:2021jqk}, among other areas.

 In particular, just as dislocations and disclinations in crystals are described by torsion and curvature, respectively, gravitational theories with torsion can support localized and topologically nontrivial configurations. These defects may be regarded as the gravitational analogues of dislocations or vortices. Within this framework, one can use the AdS/CFT correspondence \cite{Maldacena:1997re,Gubser:2002tv,Witten:1998qj} to model topologically stable dislocations in condensed matter, by mapping asymptotically AdS bulk solutions with gravitational defects to holographic lattice distortions on the boundary.

 To construct such models, it is useful to focus on gravitational theories where topological structures play a central role. A natural choice in this sense  is Einstein–Gauss–Bonnet AdS gravity at the Chern–Simons (CS) point, known as Chern–Simons AdS gravity in five dimensions, since it is topological by construction. In this theory, the action takes the form of a Chern–Simons functional for the AdS group, and solutions with nontrivial topological charge (arising from the homotopy of the gauge connection) are expected and often more natural than in Einstein gravity.

Since the role of torsion in gravity closely parallels its role in condensed matter systems, it is instructive to review how dislocations are understood in the latter. In condensed matter physics, dislocations are the crystal-lattice defects associated with the discrete translational lattice symmetry. In the continuum elastic theory dislocations correspond to a torsion field, with their topological charge, the Burgers vector, being the deficiency or excess translation on the lattice, representing a point-like source of the torsion \cite{kleinert1989gauge}. Therefore, a single dislocation may be thought of as the ``torsional vortex'' with the topological charge given by the Burgers vector. More precisely, the flux of the torsion field in the continuum elastic theory yields the Burgers vector of the dislocation defect, since the Burgers vector and torsion represent obstructions to the perfect translational periodic order in the corresponding discrete and continuum medium, respectively. These features motivate a dislocation in the bulk gravitational theory being represented in terms of the torsional field. 

Guided by these physically motivated principles, we construct a topological defect solution in a $(3+1)$-dimensional spacetime, arising from the bulk  $(4+1)$-dimensional Chern–Simons gravitational theory in asymptotically AdS spacetime, with the black hole geometry encoding the temperature and torsion representing the dislocation defects. We explicitly solve the bulk equations of motion using an asymptotic Fefferman–Graham expansion \cite{Fefferman-Graham} near the boundary. Our analysis shows that the boundary theory admits a family of axially symmetric solutions localized on a (purely spatial) ring of finite radius, as given in Eq.~\eqref{rho'}. For such configurations, we identify a possible nontrivial Abelian anomaly that scales with the Nieh–Yan invariant \cite{Nieh:1981ww}, explicitly expressed in Eq.~\eqref{eq:NY-anomaly-solution}. However, establishing its precise relation to the chiral anomaly requires further investigation.

The rest of the paper is organized as follows. In Sec.~\ref{Sec:EGB},  we discuss Einstein-Gauss-Bonnet gravity in $4+1$ spacetime dimensions from which the gravitational CS theory emerges. The holographic theory is introduced in Sec.~\ref{Sec:HFT}, with temperature encoded through a black hole and dislocations represented by a torsion field. In Sec.~\ref{Sec:Holographic-WSM}, we show that purely axial torsion does not admit a nontrivial solution, while in Sec.~\ref{WSM} we explicitly find the solutions representing the holographic dislocation and compute the corresponding Burgers vector. Concluding remarks are presented in Sec.~\ref{Sec:Conclusions}. Details of notation and calculations are relegated to several Appendices.

\section{Einstein-Gauss-Bonnet  \texorpdfstring{AdS$_5$}{AdS5} gravity}
\label{Sec:EGB}

Consider five-dimensional Einstein-Gauss-Bonnet (EGB) gravity
\cite{Boulware:1985wk} with negative cosmological constant $\Lambda <0$,
described by the action functional%
\begin{equation}
I_{\mathrm{EGB}}[\hat{g}]=\frac{1}{16\pi G}\int \mathrm{d}^{5}x\,\sqrt{-\hat{%
g}}\left( \hat{R}-2\Lambda +\alpha \,\mathfrak{R}^{2}\right) \,,
\end{equation}%
where $G$ is the gravitational constant of dimension $G\sim $ (length)$^3$ in the natural units, $\hat{g}_{MN}(x)$, $M,N=0,\ldots 4$, is the space-time metric with the mostly positive signature and $\mathfrak{R}^{2}$ denotes the Gauss-Bonnet term quadratic in the Riemann tensor\footnote{The Gauss-Bonnet term written in terms of the Riemann tensor $\hat{R}_{\ NKL}^{M}$, Ricci tensor $\hat{R}_{MN}=\hat{R}_{\ MKN}^{K}$, and
Ricci scalar $\hat{R}=\hat{g}^{MN}\hat{R}_{MN}$, has the form $\Re ^{2}=%
\hat{R}^{MNKL}\hat{R}_{MNKL}-4\hat{R}^{MN}\hat{R}_{MN}+\hat{R}^{2}$.}, with the associated coupling constant $\alpha \sim \text{(length)}^2$. It is the most general higher-curvature correction of General Relativity in $4+1$ dimensions still possessing second-order gravitational equations. 

In the first-order formulation, the above action can be expressed in terms of the vielbein 1-form $\hat{e}^{A}=\hat{e}_{M}^{A}(x)\,\mathrm{d}x^{M}
$,  using the differential forms in the local
coordinate basis $\mathrm{d}x^{M}$, and $A=0,\ldots 4$ are Lorentz indices of the tangent space. The vielbein and its inverse project the bulk tensors to the tangent
tensors, and \textit{vice versa}. The flat tangent space is endowed with the
Minkowski metric $\eta _{AB}=\mathrm{diag}\left( -,+,+,+,+\right) $. In this formalism, the fundamental field is the vielbein instead of the
metric, which can be obtained as $\hat{g}_{MN}=\eta _{AB}\,\hat{e}_{M}^{A}\hat{e}_{N}^{B}$.

An advantage of working in the first-order framework is a natural introduction
of the torsion field, necessary for sourcing fermionic fields. Namely, in the Riemann space (without torsion), the parallel transport
is defined by the Christoffel symbol, which translates here to the parallel transport of Lorentz vectors performed by the
spin connection 1-form $\hat{\omega}^{AB}=\hat{\omega}_{M}^{AB}(x)\,\mathrm{d
}x^{M}$, which is determined by  $\hat e$ up to gauge transformations. In the Riemann-Cartan space (with torsion), the spin-connection becomes an independent gauge field. The associated Lorentz field-strength 2-form is $\hat{R}^{AB}=\mathrm{d}\hat{\omega}^{AB}+\hat{\omega}_{\ C}^{A}\wedge \hat{\omega}^{CB}$. The Riemann tensor is obtained
from $\hat{R}^{AB}=\frac{1}{2}\,\hat{R}_{MN}^{AB}\,\mathrm{d}x^{M}\wedge
\mathrm{d}x^{N}$ by projecting the Lorentz indices of the Lorentz
field-strength to the spacetime indices as $\hat{R}_{MN}^{KL}=\hat{R}%
_{MN}^{AB}\,\hat{e}_{A}^{K}\hat{e}_{B}^{L}$, using the inverse vielbein $%
\hat{e}_{A}^{M}$.

With the above definitions at hand, and skipping the wedge between the forms for the sake of simplicity, the EGB AdS action acquires the form 
\begin{equation}
I_{\mathrm{EGB}}[\hat{e},\hat{\omega}]=\kappa \int \epsilon _{ABCDE}\left(
\frac{4\alpha }{\ell ^{3}}\,\hat{R}^{AB}\hat{R}^{CD}+\frac{2}{3\ell ^{3}}\, 
\hat{R}^{AB}\hat{e}^{C}\hat{e}^{D}+\frac{1}{5\ell ^{5}}\,\hat{e}^{A}\hat{e}
^{B}\hat{e}^{C}\hat{e}^{D}\right) \hat{e}^{E}\,, \label{EGB}
\end{equation}
where we redefined the dimensionless gravitational constant as $\kappa =\frac{\ell ^{3}}{64\pi G}$, and introduced $\ell ^{2}=-\frac{3}{\Lambda }$, where $\ell $ has
dimensions of the radius. We use the notation of Appendix \ref{conventions} for the Levi-Civita symbol, given by Eq.~\eqref{volume 4D}.  The gravitational constant and the radius are the only two independent coupling constants in the theory.

A space of solutions of $I_{\mathrm{EGB}}[\hat{g}]$ coincides with the space
of solutions of $I_{\mathrm{EGB}}[\hat{e},\hat{\omega}]$ only when one imposes that the spacetime torsion vanishes, $\hat{T}^{A}=\hat{\mathrm{D}}e^{A}=\mathrm{d}\hat{e}^{A}+\hat{\omega}^{AB}\wedge \hat{e}_{B}$=0, where $\hat{\mathrm{D}}$ is the Lorentz covariant derivative. When $\hat{T}^{A}\neq 0$, the EGB gravity  \eqref{EGB} has more degrees of freedom than General Relativity because the spin connection is dynamical.

\paragraph{Chern-Simons AdS gravity.}

For holographic applications, we need the spacetime asymptotically approaching AdS space.
Thus, we need the theory to admit a global AdS solution, which will play a role in the ground state of the theory. However, due to the quadratic term in $\hat{R}^{AB}$ in the action, there are two global AdS solutions for
generic values of $\alpha >\alpha _{\mathrm{CS}}$, and only one solution in
the so-called Chern-Simons point $\alpha _{\mathrm{CS}}=\frac{\ell ^{2}}{4}$. 

We are interested in EGB in the Chern-Simons point when the EGB theory becomes CS AdS gravity,
\begin{equation}
I_{\mathrm{CS}}[\hat{e},\hat{\omega}]=\kappa \int \epsilon _{ABCDE}\left(
\frac{1}{\ell }\,\hat{R}^{AB}\hat{R}^{CD}+\frac{2}{3\ell ^{3}}\,\hat{R}^{AB}
\hat{e}^{C}\hat{e}^{D}+\frac{1}{5\ell ^{5}}\,\hat{e}^{A}\hat{e}^{B}\hat{e}%
^{C}\hat{e}^{D}\right) \hat{e}^{E}. \label{CS}
\end{equation}
The unique global AdS vacuum of CS AdS gravity \eqref{CS} is given by the solution $\hat{R}^{AB}=-\frac{1}{\ell ^{2}}\,\hat{e}^{A}\wedge \hat{e}^{B}$, therefore $\ell $  
can be identified as the AdS
radius. Note that the AdS radius in CS gravity does not coincide with the
AdS radius in Einstein-Hilbert (EH) gravity ($\alpha =0$, $\Lambda_{\mathrm{EH}}=-\frac{6}{\ell_{\mathrm{EH}}^{2}}$).

The limit $\alpha \rightarrow \alpha _{\mathrm{CS}}$ is discontinuous
because two theories differ drastically. While EH and EGB are gauge
invariant only under Lorentz symmetry, the CS gravity with negative cosmological constant becomes a gauge theory for the AdS group.

This can be seen explicitly by rewriting the CS action in terms of a
dynamical Lie-algebra valued gauge field 1-form $A=A_{M}(x)\,\mathrm{d}x^{M}$
that is a connection on a fiber bundle,%
\begin{equation}
A=\frac{1}{\ell }\,\hat{e}^{A}P_{A}+\frac{1}{2}\,\hat{\omega}^{AB}J_{AB}\,,
\label{A.comp}
\end{equation}
where $\{ J_{AB}=-J_{BA},P_{A} \}$ are the generators of $SO(4,2)$,
locally isomorphic with AdS$_{5}$ group. For the explicit algebra, see Eq.~\eqref{AdS algebra 1} in Appendix \ref{conventions}. The associated Lie-algebra valued
field strength 2-form is $F=\frac{1}{2}\,F_{MN}(x)\,\mathrm{d}x^{M}\wedge
\mathrm{d}x^{N}=\mathrm{d}A+A\wedge A$, with the components%
\begin{equation}
F=\frac{1}{\ell }\,\hat{T}^{A}P_{A}+\frac{1}{2}\,\left( \hat{R}^{AB}+\frac{1%
}{\ell ^{2}}\,\hat{e}^{A}\wedge \hat{e}^{B}\right) J_{AB}\,.
\end{equation}%
Therefore, $F^{AB}=\hat{R}^{AB}+\frac{1}{\ell ^{2}}\,\hat{e}^{A}\wedge \hat{e}^{B}$ is the AdS curvature of the spacetime, with the AdS vacuum that is
the pure gauge ($F^{AB}=0$) of the theory.

For completeness, we write the CS AdS gravity as a gauge theory for AdS$_{5}$
group explicitly,
\begin{equation}
I_{\mathrm{CS}}[A]=\frac{\kappa }{3}\int \mathrm{Tr}\left( AF^{2}-\frac{1}{2}%
\,A^{3}F+\frac{1}{10}\,A^{5}\right) \,,  \label{CS action}
\end{equation}%
where $\kappa$ is the level of the CS action and the trace is defined as a symmetric invariant tensor of the AdS group,
with the only non-vanishing components
\begin{equation}
\frac{1}{4}\, \mathrm{Tr}\left( J_{AB}J_{CD}P_{E}\right) =\epsilon _{ABCDE}\,.
\end{equation}
We neglected all the boundary terms. The action principle problem and the counterterms that render the action \eqref{CS action} finite have been analyzed in \cite{Banados:2006fe}. For a comprehensive review on CS (super)gravities, see Ref.~\cite{Zanelli:2005sa}.

Varying the action functional \eqref{CS action} and asking it to be stationary, we arrive at the equations of motion
\begin{eqnarray}
\delta \hat{e}^{E} &:&\quad 0= \epsilon _{ABCDE}\,F^{AB}\wedge F^{CD}\,,  \notag \\
\delta \hat{\omega}^{DE} &:&\quad 0=\epsilon _{ABCDE}\,F^{AB}\wedge \hat{T}^{C}\,.
\label{eom}
\end{eqnarray}
The first is the generalized EGB field equation that includes torsional degrees of freedom, while the second is characteristic of the $\hat T^A \neq 0$ sector. We are interested in a non-trivial torsion field, as it is a source for fermions in a holographically dual theory. 

Solutions of five-dimensional CS (super)gravity with non-vanishing torsion have been previously considered in the literature. A stable global AdS$_{5}$ geometry containing Abelian matter with nontrivial winding was discussed in
\cite{Miskovic:2006ei}, while degenerate AdS$_{5}$ black holes were
investigated in \cite{Aros:2006qc}. More in line with our work, CS
AdS$_{5}$ black holes with axial torsion were analyzed in \cite{Canfora:2007xs} and with the addition of Abelian and non-Abelian solitons in \cite{Andrianopoli:2021qli}. Other black hole solutions in CS AdS$_{5}$
gravity were discussed in \cite{Giribet:2014hpa}. When the cosmological
constant is slightly modified,  CS gravity becomes effectively EGB gravity,
which was studied \cite{Banados:2001hm}.

The action \eqref{CS action} and the equations of motion \eqref{eom} describing the bulk dynamics of the gravitational fields $\hat e^A_M(x)$ and $\hat \omega^{AB}_M(x)$ are our starting point for constructing the holographic theory.

\section{Holographic field theory}
\label{Sec:HFT}

The holographic dictionary provides an equivalence between a classical gravity in AdS space, and a strongly-coupled quantum field theory (QFT) on the boundary in one dimension less \cite{Maldacena:1997re,Gubser:2002tv,Witten:1998qj}. This duality is given through the equality of quantum partition functions in two theories. Applied to our case, in the gravity side, we have the CS partition function in classical approximation,
\begin{equation}
Z_{\mathrm{CS}}[A_{(0)}]\simeq \left. \mathrm{e}^{\mathrm{i}I_{\mathrm{CS}}^{\mathrm{ren}}[A]}\right\vert _{A_{(0)}}\,,
\label{Zcs}
\end{equation}
where $I_{\mathrm{CS}}^{\mathrm{ren}}=I_{\mathrm{CS}}+I_{\mathrm{B}}$ is the renormalized
CS action that includes a surface term $I_{\mathrm{B}}$ that ensures that the action
principle is satisfied upon suitable boundary conditions on the gauge field  $A \to A_{(0)}$, and it also contains counterterms that make $I_{\mathrm{CS}}^{\mathrm{ren}}$ finite in the asymptotic region.
As known, AdS space has large-distance divergences because the metric on the boundary has a pole of order two \cite{Skenderis:2002wp}.

On the other hand, holographically dual conformal field theory (CFT), that is, QFT, has the partition function given by
\begin{equation}
Z_{\mathrm{QFT}}[A_{(0)}]=\mathrm{e}^{\mathrm{i}W[A_{(0)}]}\,,
\end{equation}%
where $W[A_{(0)}]$ is the (non-local) quantum effective action where all the
boundary fields, coupled to the external source $A_{(0)}$, have been
integrated out. Then, the AdS/CFT correspondence equals the renormalized
classical AdS gravity action  in five dimensions, with the  quantum effective action in four dimensions,
\begin{equation}
I_{\mathrm{CS}}^{\mathrm{ren}}[A_{(0)}]\simeq W[A_{(0)}]\,. \label{eff}
\end{equation}

Since the on-shell action $I_{\mathrm{CS}}^{\mathrm{ren}}$ is a boundary
term, it is defined in one dimension less, where all holographic observables
become functions of the boundary coordinate $x^{\mu }$. In that way, the above holographic identification enables the computation of the $n$-point functions in CFT directly from the renormalized gravitational action.

In particular, we are interested in the 1-point functions, where the
conserved currents $\left\langle J^{\mu }\right\rangle $ in CFT are obtained
from the variation of the sources $A_{(0)\mu }$ as
\begin{equation}
\delta W[A_{(0)}]=\int \mathrm{d}^{4}x\,\delta A_{(0)\mu }\left\langle J^{\mu
}\right\rangle \quad \Rightarrow \quad \left\langle J^{\mu }(x)\right\rangle
=\frac{\delta W[A_{(0)}]}{\delta A_{(0)\mu }(x)}\simeq \frac{\delta I_{\mathrm{CS}
}^{\mathrm{ren}}[A_{(0)}]}{\delta A_{(0)\mu }(x)}\,. 
\label{dW}
\end{equation}
Since the on-shell bulk action $I_{\mathrm{CS}}^{\mathrm{ren}}[A_{(0)}]$ is known, the expression for
$\left\langle J^{\mu }(x)\right\rangle $ can be computed, as well. If the Lie-algebra valued current satisfies the classical conservation law $D_{(0)\mu} J^\mu=0$, then the quantum Ward identity 
\begin{equation}\label{eq:general-anomaly}
D _{(0)\mu}\left\langle J^{\mu }(x)\right\rangle =\mathcal{A}(x)\,,
\end{equation}
can be obtained straightforwardly by taking a covariant derivative with respect to $A_{(0)}$, exhibiting a quantum anomaly if $\mathcal{A}$ is non-trivial. In that case, the anomaly is generated by the background field $F_{(0)\mu\nu}$.

Let us point out that all classical computations and on-shell expressions
obtained in the bulk are mapped, via the AdS/CFT correspondence, into the quantum off-shell expressions in the boundary theory.

\subsection{Our setup}
\label{setup}

We apply the above mechanism to a setting where AdS symmetry is broken, focusing on a specific black hole solution in asymptotically AdS$_5$ spacetime. As a result, the conformal symmetry on the boundary is also broken, yielding a particular QFT. Since black holes introduce temperature in the holographic theory, this approach has been employed in condensed matter physics, in particular in the study of holographic superconductors (see \cite{Hartnoll:2008vx,Hartnoll:2008kx,Horowitz:2008bn}).

A key feature of the holographic prescription is that it can describe fermionic quasiparticles on the boundary without requiring fermions in the bulk theory. The effective action $W[A_{(0)}]$, defined in \eqref{eff} with fermionic fields integrated out, depends solely on bosonic sources. Nonetheless, the resulting 
$n$-point functions fully capture the quantum properties of the boundary fermionic theory.

In our case, the bulk field given by \eqref{A.comp} has gravitational components $\hat{e}_{M}^{A}$ and $\hat{\omega}_{M}^{AB}$, and the corresponding
boundary sources are $e_{\mu }^{a}$ and $\omega _{\mu }^{ab}$, respectively. The variation \eqref{dW} in that case
reads
\begin{equation}
\delta W[e,\omega ]=\int \mathrm{d}^{4}x\,|e|\left( \delta e_{\mu
}^{a}\,\left\langle \tau _{~ a}^{\mu }\right\rangle +\frac{1}{2}\,\delta\omega _{\mu
}^{ab}\,\left\langle \sigma _{~ ab}^{\mu }\right\rangle \right) ,
\label{delta W}
\end{equation}
where $e=\det [e_{\mu }^{a}]$ is the Jacobian, $\tau _{~ a}^{\mu }$ is the energy-momentum
tensor of the holographic theory and $\sigma _{~ ab}^{\mu }$ is the spin current.  They
describe conserved currents of a Poincar\'{e}-invariant action of 
fermionic fields in curved background, that is, fermionic fields coupled
to the external sources $e_{\mu }^{a}$ and $ \omega _{\mu }^{ab}$. The
classical conservation laws of these currents associated with the diffeomorphisms
and Lorentz transformations are, respectively,
\begin{eqnarray}
D_{\mu }\left( |e|\tau _{~ a}^{\mu }\right) e_{\nu }^{a} &=&|e|(\tau _{~ a}^{\mu
}T_{\mu \nu }^{a}+\frac{1}{2}\,\sigma _{~ ab}^{\mu }R_{\mu \nu }^{ab})\,,
\notag \\
D_{\mu }\left( |e|\sigma _{~ ab}^{\mu }\right) \  &=&|e|(\tau _{ab}-\tau
_{ba})\,,
\label{conservation}
\end{eqnarray}
where $\tau _{ab}=e_{\mu a}\tau _{~b}^{\mu }$ and $D_{\mu }(\omega )$ is a
Lorentz-covariant derivative. In the Riemann space, both $\sigma _{~  ab}^{\mu }$ and $T^a_{\mu\nu}$ vanish, and the stress tensor $\tau _{~ a}^{\mu }$ becomes symmetric and covariantly conserved, as expected.

In a Riemann-Cartan background, one can introduce an odd-parity Abelian current defined as the completely antisymmetric component of the spin current,
\begin{equation} J_{\mathrm{odd}}^{\mu }=\frac{1}{3!}\,\epsilon ^{abcd}\,e_{a}^{\mu }\sigma_{bcd}\,, \label{J} \end{equation}
with $\sigma _{bcd}=e_{b\mu }\sigma _{~cd}^{\mu }$. As argued in \cite{Obukhov:1982da,Obukhov:1983mm,Urrutia-Vergara}, this current coincides with the chiral current, $J_{\mathrm{ch}}^{\mu }$, characteristic of Weyl fermions. In fact, for a four-dimensional Weyl fermion on a Riemann–Cartan background, described by
$\mathcal{L}_{\mathrm{W}}=\mathrm{i}\sqrt{g}\,\overline{\psi }\gamma ^{\mu }(\partial
_{\mu }+\frac{1}{2}\,\omega ^{ab}\Sigma _{ab})\psi $,
this is indeed satisfied, as $J_{\mathrm{odd}}^{\mu }=J_{\mathrm{ch}}^{\mu }\equiv \overline{\psi }\gamma^{\mu }\gamma _{5}\psi$ holds.

It is well known that the chiral current is classically conserved but acquires a quantum anomaly, ${\mathcal{A}}_{\mathrm{ch}}$, when a chiral fermion couples to an external electromagnetic field. In flat spacetime, this anomaly is proportional to the Pontryagin density, $\frac{1}{8}\,\epsilon ^{\mu \nu \alpha \beta }\,F_{\mu \nu }F_{\alpha \beta }=\mathbf{E}\cdot \mathbf{B}$, where $\mathbf{E}$ and $\mathbf{B}$ are the electric and magnetic fields. This mechanism has been widely used to describe the chiral anomaly in Weyl semimetals (for a review, see Ref.~\cite{Burkov-review-WSM}).

In the present holographic theory, however, it is not yet clear whether $J_{\mathrm{odd}}^\mu$ is conserved at the classical level and anomalous in the quantum theory. If so, whether the corresponding anomaly
\begin{equation} 
\partial _{\mu }\left( \left\vert e\right\vert \left\langle J_{\mathrm{odd}}^{\mu }\right\rangle \right) =|e|{\mathcal{A}}_{\mathrm{odd}}\,, \label{odd.anomaly} 
\end{equation} 
should be interpreted as a chiral anomaly sourced by the torsion field. This is a question that can be addressed using holography.

To construct holographic QFTs, we begin with a bulk theory based on Einstein–Gauss–Bonnet AdS gravity, which is known to describe the normal phase of holographic superconductors \cite{Gregory:2009fj,Barclay:2010up,Pan:2009xa}. By incorporating torsional degrees of freedom, we enable the background to support dislocations known to produce an imbalance in left- and right-handed chiral quasiparticles. Finally, since Chern–Simons terms universally generate anomalies, we fix the Gauss–Bonnet coupling to its CS value, thereby identifying CS AdS gravity as our bulk gravitational theory.\medskip

CS holography has been developed in \cite{Banados:2006fe} in a general non-Abelian CS theory and CS AdS gravity case with arbitrary sources. The goal of this work is to apply these results to the particular sources corresponding to the black hole geometry in the bulk and find a spin connection that gives rise to  topological defects or dislocations in a boundary QFT.

\subsection{Obtaining holographic equations}
\label{Obtaining Hol Eqs}

A holographic theory is identified with the asymptotic sector of CS AdS gravity. To explore it, we have to solve part of the gravitational field equations determining a radial evolution. The remaining ones are the holographic equations that describe the properties of the asymptotic theory. This analysis is performed in Ref.~\cite{Banados:2006fe}, and we start with a review of the results.

To take the asymptotic limit of CS gravity and analyze the resulting boundary theory, first-order differential equations \eqref{eom} in the radial coordinate have to be solved for given boundary conditions, determining the radial evolution of the fields. 

Let us choose the local
coordinates at the 5D spacetime manifold $\mathcal{M}=\mathbb{R}\times
\Sigma $, with the spatial section $\Sigma $, as $x^{M}=(x^{\mu },\sigma )$,
where $\sigma \geq 0$ is the radial coordinate, and $x^{\mu }$, $\mu =0,1,2,3
$, parameterizes the 4D asymptotic boundary $\partial \mathcal{M}=\mathbb{R}
\times \partial \Sigma _{\infty }$, located at $\sigma =\sigma _{\mathrm{B}}=
\mathrm{const}$. It is convenient to choose $\sigma _{\mathrm{B}}=0$ because
we are interested in a near-boundary analysis. The group indices decompose
accordingly as $A=(a,4)$. 

Starting from this section, we will also set the AdS radius to one for simplicity, $\ell =1$. Note that we do not lose generality with this choice
because the holographic prescription that we use is valid only in AdS spaces.
Furthermore, the flat limit $\ell \rightarrow \infty $ is not well-defined;
for instance, the counterterms used to regularize CS AdS gravity are
proportional to $\ell $ \cite{Miskovic:2007mg}.

We impose the following 15 holographic gauge-fixing conditions of the 15-dimensional $\mathrm{SO}(4,2)$ gauge symmetry \cite{Banados:2006fe}, 
\begin{equation}
A_{\sigma }=-\frac{1}{2\sigma }\,P_{4}\,\quad \Leftrightarrow \quad \hat{e}_{\sigma }^{4}=-\frac{1}{2\sigma }\,,\quad \hat{e}_{\sigma }^{a}=0\,,\quad \hat{\omega}_{\sigma }^{AB}=0\,.
\end{equation}
The component $\hat{e}_{\sigma }^{4}$ is chosen such that the vielbein is
invertible, and $\sigma $ becomes the dimensionless Fefferman-Graham (FG) radial coordinate \cite{Fefferman-Graham}. An additional useful condition that makes the boundary orthogonal to the radial coordinate is
\begin{equation}
\hat{e}_{\mu }^{4}=0\,.
\end{equation}
It can be shown that the above condition is always allowed because it fixes the residual gauge symmetry. In that way, $A_{\sigma }$ is determined by the gauge choice, and $A_{\mu }$ can be
obtained dynamically, as the exact solution of the subset of equations \eqref{eom} that contain radial derivatives,
\begin{equation}
A_{\mu }(x,\sigma )=\sigma ^{\frac{1}{2}\, P_{4}}\,A_{\mu }(x,0)\,\sigma ^{-\frac{1}{2}\,P_{4}}. \label{A(s,x)}
\end{equation}
The boundary value of the gauge field $A_{(0)}(x)$ that appears in the gravitational partition function \eqref{Zcs} is the boundary 1-form $A(x,0)=A_{\mu }(x,0)\,\mathrm{d}x^{\mu }$. It can be expanded in the AdS basis $J_{a}^{\pm }=P_{a}\pm J_{a4}$ (for the algebra, see \eqref{AdS algebra 2} in Appendix \ref{conventions}) as
\begin{equation}
\partial \mathcal{M}: \quad A(x,0)=A_{\mu }(x,0)\,\mathrm{d}x^{\mu }= e^{a}(x)J_{a}^{+}+\,k_{\mu }^{a}(x)J_{a}^{-}+\frac{1}{2}\,\omega ^{ab}(x)J_{ab}\,. \label{A(0,x)}
\end{equation}

Combining \eqref{A(s,x)} and \eqref{A(0,x)} and using a few algebraic identities\footnote{The identities used are $\sigma ^{\frac{1}{2}\,P_{4}}\,J_{a}^{\pm }\,\sigma ^{-\frac{1}{2}\,P_{4}}=\sigma^{\mp \frac{1}{2}}\,J_{a}^{\pm }$ and $\sigma ^{\frac{1}{2}\,P_{4}}\,J_{ab}\,\sigma ^{-\frac{1}{2}\,P_{4}}=J_{ab}$.}, the radial
evolution of the gauge field acquires the form
\begin{equation}
A(x,\sigma )=\frac{1}{ \sqrt{\sigma }}\,e^{a}(x)\,J_{a}^{+}+\sqrt{\sigma }k^{a}(x)\,J_{a}^{-}+\frac{1}{2}\,\omega
^{ab}(x)\,J_{ab}\,.
\end{equation}
It means that the bulk vielbein and the spin connection have radial dependence given by
\begin{align}
\hat{e}^{a}& =\frac{1}{\sqrt{\sigma }}\,e^{a}+\sqrt{\sigma }\,k^{a}\,,
\notag \\
\hat{\omega}^{ab}& =\omega ^{ab}\,,  \label{expansion} \\
\hat{\omega}^{a4}& =\frac{1}{ \sqrt{\sigma }}\,e^{a}-\sqrt{\sigma }k^{a}\,.  \notag
\end{align}

 When the radial expansion of $\hat{e}^{A}$ is 
 inserted
 in the five-dimensional line element $d\hat{s}^{2}=\eta _{AB}\hat{e}^{A}\hat{e}^{B}$,  it acquires the standard FG form \cite{Fefferman-Graham},
\begin{equation}
\mathrm{d}s^{2}=\frac{\mathrm{d}\sigma ^{2}}{4\sigma ^{2}}+\frac{1}{\sigma }\,\left[ g_{\mu \nu }+\sigma \left( k_{\mu \nu
}+k_{\nu \mu }\right) +\sigma ^{2}\,k_{a\mu }\,k_{\ \nu }^{b}\right]\,\mathrm{d}x^{\mu }\mathrm{d}x^{\nu },  \label{FGmetric}
\end{equation}
where the induced metric on the surface $\sigma =\mathrm{const.}$ expands in a way consistent with asymptotically AdS spaces, that is, $\frac{1}{\sigma }\times\,$(regular part) when $\sigma\to 0$.

We denoted the boundary metric source  by $g_{\mu \nu }(x)=\eta _{ab}e_{\mu
}^{a}e_{\nu }^{b}$, which is the leading order in the expansion of the induced metric. Similarly, we can recognize the holographic sources in the first-order formulation as the leading orders in the near-boundary expansion \eqref{expansion}. Namely, when $\sigma \rightarrow 0$ then, up to a
conformal factor, the boundary conditions of the fields are $\hat{e}^{a}\rightarrow e^{a}$ and $\hat{\omega}^{ab}\rightarrow \omega ^{ab}$.
Thus, $e^{a}$ and $\omega ^{ab}$ are the boundary sources.

As a consequence of the radial expansion \eqref{expansion}, the AdS field strength expands as 
\begin{align}
F^{ab}& =R^{ab}+2\left( e^{a}\wedge k^{b}-e^{b}\wedge
k^{a}\right) \,,  \notag \\
F^{a4}& =\frac{1}{\sqrt{\sigma }}\ T^{a}-\sqrt{\sigma }\,
\mathrm{D}k^{a}\ ,  \notag \\
\hat{T}^{a}& =\frac{1}{\sqrt{\sigma }}\,T^{a}+\sqrt{\sigma }\mathrm{D}%
k^{a}\,,  \label{Fexp} \\
\hat{T}^{4}& =-2\,e^{a}\wedge k_{a}\,,  \notag 
\end{align}
where $\mathrm{D}=\de x^\mu \mathrm{D}_\mu $ is the Lorentz-covariant derivative with respect to the connection $\omega ^{ab}$ on the boundary and $T^{a}=\mathrm{D}e^{a}$. Also, $R^{ab}=\mathrm{d}\omega ^{ab}+\omega ^{ac}\wedge \omega _{c}^{\ b}$ is the
curvature $2$-form on the boundary, and we use the notation $\epsilon _{abcd4}=\epsilon _{abcd}$ (see \eqref{Levi 3D} in notation summary in Appendix \ref{conventions}). 

Finally, the equations of motion \eqref{eom} that do not contain radial derivatives should be used to determine all the boundary fields in terms of
the sources. These constraints, important because they determine the dynamics of the boundary, read
\begin{align}
C\;& =\epsilon _{abcd}\, F^{ab} \wedge F^{cd} =0\,,  \notag \\
C_{a}& = \epsilon _{abcd}\,F^{bc} \wedge T^{d}=0\,,  \notag \\
\bar{C}_{a}& = \epsilon _{abcd}\,F^{bc}\wedge \mathrm{D}k^{d}=0\,,  \label{constraints} \\
C_{ab}& = \epsilon _{abcd}\,\left( F^{cd} \wedge e^{e}\wedge k_{e}+2T^{c}\wedge \mathrm{D}
k^{d}\right)=0 \,.  \notag
\end{align}
Note that they do not depend on the radial coordinate $\sigma$, so they are truly holographic equations on the boundary. They govern the dynamics of the boundary fields  $e^{a}(x)$, $\omega^{ab}(x)$ and $k^{a}(x)$, thus will call them the holographic equations.

\subsection{Introducing temperature}
\label{Temperature}

To identify the holographic field theory as a thermal system, it has to be placed at a constant temperature $T$. Thus, the gravitational dual has to be a black hole, with the Hawking temperature $T$.

It is known that the CS AdS gravity possesses a static, spherically
symmetric black hole solution. In the Schwarzschild-like coordinates $(t,y^{i},r)$, where $r$ is the radial coordinate with the asymptotic boundary at $r\rightarrow \infty $, it has the form \cite{Banados:1993ur}
\begin{equation}
\de s^{2}=-f^{2}(r)\de t^{2}+\frac{\de r^{2}}{f^{2}(r)}+r^{2}\de\Omega
^{2}\,,  \label{BHmetric}
\end{equation}
where $f(r)=\sqrt{r^{2}-M}$ is the metric function.
Because this metric is the dimensional continuation of the three-dimensional BTZ black hole geometry \cite{Banados:1992wn}, the above solution is also called dimensionally
continued black hole.

In Eq.~\eqref{BHmetric}, $\de\Omega $ is a line element of the three-dimensional transverse
submanifold $t,r=\mathrm{const}$, which is the maximally symmetric space of
unit radius. If the ``angles'' in this
space are denoted by $y^{i}$ and the metric $\gamma _{ij}(y)$, then $\de\Omega
^{2}=\gamma _{ij}(y)\de y^{i}\de y^{j}$. In AdS space, the transverse space
can have curvature $\varkappa =0,1$ and $-1$, corresponding to the flat,
spherical, and hyperbolic horizon topology, respectively. Taking this into
account, we have $M=\mu-\varkappa \geq 0$, where $M$ is a non-negative parameter to ensure the existence of the horizon, and $\mu$ is the dimensionless mass parameter of the black hole.

The black hole \eqref{BHmetric} has the horizon
\begin{equation}
f^{2}(r_{\mathrm{H}})=0\quad \Rightarrow \quad r_{\mathrm{H}}= \sqrt{M}\,,\quad M=\mu -\varkappa \geq 0\,,  \label{horizon}
\end{equation}
with the Hawking temperature
\begin{equation}
T=\left. \frac{(f^2)'}{4\pi }\right\vert _{r_{\mathrm{H}}}=\frac{\sqrt{M}}{2\pi } = \frac{r_{\mathrm{H}}}{2\pi } \,.
\label{T}
\end{equation}
The temperature proportional to the horizon radius is typical of dimensionally continued black holes.

A relation between the FG radial coordinate $\sigma $ given by 
\eqref{FGmetric} with the radial coordinate $r$ in the black hole 
\eqref{BHmetric} is given by\footnote{In the derivation, we used the integral $\int
\frac{\mathrm{d}x}{\sqrt{x^{2}-M}}=\ln (x+\sqrt{x^{2}-M})$.}
\begin{equation}
\frac{\de r^{2}}{f^{2}(r)}=\frac{\de\sigma ^{2}}{4\sigma ^{2}}\quad
\Rightarrow \quad \frac{\de r}{f}=-\frac{ \de\sigma }{2\sigma }\quad
\Rightarrow \quad \sigma =\frac{1}{\left( r+f\right) ^{2}}\,,
\label{sigma}
\end{equation}
where the sign in the second step was chosen such that $\sigma \rightarrow 0$
as $r\rightarrow \infty $.  Inverting the relation, 
\begin{equation}
r=\frac{1}{2}\left( M\sqrt{\sigma }+\frac{1}{\sqrt{\sigma }}\right) .
\end{equation}
The black hole metric rewritten in terms of the FG coordinates reads
\begin{equation}
\de s^{2}=\frac{\de\sigma ^{2}}{4\sigma ^{2}}-\frac{(M\sigma -1)^{2}%
}{4\sigma }\,\de t^{2}+\frac{(M\sigma +1)^{2}}{4\sigma }\,\gamma
^{ij}(y)\de y^{i}\de y^{j}.
\end{equation}
To write the corresponding vielbein, let us decompose the 5D Lorentz indices as $a=(0,i,4)$ and\ denote the
three-dimensional vielbein by $\tilde{e}^{m}=\tilde{e}_{i}^{m}(y)\,
\mathrm{d}y^{i}$,  such that $\gamma _{ij}=\delta _{mn}\tilde{e}_{i}^{m}
\tilde{e}_{j}^{n}$. Then, the non-zero components of the five-dimensional vielbein read
\begin{equation}
\hat{e}^{0}=\frac{1-M\sigma }{2\sqrt{\sigma }}\,\mathrm{d}t\,,\qquad \hat{e}^{m}=\dfrac{ 1+M\sigma}{2\sqrt{\sigma }}\,\tilde{e}^{m}\,,\qquad \hat{e}^{4}=-\frac{\mathrm{d}\sigma}{2\sigma } \,.  \label{e}
\end{equation}
The signs are fixed such that $\hat{e}^{0}$ and $\hat{e}^{m}$ have positive
orientations with respect to the coordinates when $M=0$, while the negative sign in $\hat{e}^{4}$ is consistent with the second expression in 
\eqref{sigma}.

To identify the four-dimensional vielbein $e^{a}$ and the quantity $k^{a}$,
we compare the induced vielbein \eqref{e} on the surface $\sigma =\mathrm{%
const}.$ with \eqref{expansion}, to obtain%
\begin{equation}
\begin{array}{llll}
e^{0} & =\dfrac{1}{2}\,\de t\,,\qquad \medskip  & k^{0} & =-\dfrac{M}{2}\,\de
t\,, \\
e^{m} & =\dfrac{1}{2}\,\tilde{e}^{m}\,, & k^{m} & =\dfrac{ M}{2}\,%
\tilde{e}^{m}\,.%
\end{array}
\label{e,k BH}
\end{equation}
Finally, we can construct the 4D metric $g_{\mu \nu }=e_{\mu }^{a}e_{a\nu }$
and the tensor $k_{\mu \nu }=e_{a\mu }k_{\ \nu }^{a}$ as%
\begin{equation}
g_{\mu \nu }=\frac{1}{4}\,\left(
\begin{array}{cc}
-1 & 0 \\
0 & \gamma _{ij}(y)
\end{array}
\right) ,\quad k_{\mu \nu }=\frac{M}{4}\,\left(
\begin{array}{cc}
1 & 0 \\
0 & \gamma _{ij}(y)
\end{array}
\right) \,,\quad \sqrt{|g|}=\frac{\sqrt{\gamma }}{16}\,,
\label{g,k}
\end{equation}
defined in a conformally flat four-dimensional space. Here, $\gamma=\det [\gamma_{ij}]$.

Until now, we obtained $e^a$ and $k^a$. However, we cannot identify the spin-connection source $\omega^{ab}$ yet because we do not have the solution for the torsion field that satisfies Eqs.~\eqref{constraints}.  Using the boundary metric \eqref{g,k}, we can find only the torsion-free spin connection $\mathring{\omega}^{ab}=0$ that satisfies $\mathring{\mathrm{D}}e^a=\de e^a+\mathring{\omega}^{ab} \wedge e_b=0$, which has only one non-zero component,
\begin{equation}
\mathring{\omega}^{12}=-\mathrm{d}\varphi \quad \Rightarrow \quad \mathring{R}^{ab}=0\,.  \label{omega}
\end{equation}
The full spin connection will be determined and discussed in Sec.~\ref{WSM}.

\subsection{Introducing dislocation}
\label{Dislocation}

In the holographic field theory, the spin connection 1-form
\begin{equation}
\omega ^{ab}=\mathring{\omega}^{ab}(e)+K^{ab}\,,\qquad \mathrm{\mathring{D}}%
e^{a}=0\,, \label{4spin}
\end{equation}%
still has to be resolved. $\mathring{\omega}^{ab}$ is the torsion-free
or Levi-Civita spin connection that is fully determined by the metric, and the
metric-independent part, $K^{ab}=-K^{ba}=K_{\ \  \mu }^{ab}\,\mathrm{d}x^{\mu }$ is  the contorsion tensor. It carries the information about the torsional degrees of freedom, with the four-dimensional torsion 2-form field defined by
\begin{equation}
T^{a}=\frac{1}{2}\,T_{\ \mu \nu }^{a}\,\mathrm{d}x^{\mu }\wedge \mathrm{d}%
x^{\nu }=\mathrm{D}e^{a}\,.
\end{equation}
There is a one-to-one relationship between the torsion and the contorsion,
\begin{equation}
T^{a}=K^{ab}\wedge e_{b}\,,
\end{equation}
or in components $T_{\mu \alpha \beta }=e_{a\mu }\,T_{\ \alpha \beta }^{a}$
and $K_{\mu \nu \alpha }=e_{a\mu }e_{b\nu }\,K_{\ \ \alpha }^{ab}$, the
relation is
\begin{equation}
T_{\mu \alpha \beta }=K_{\mu \beta \alpha }-K_{\mu \alpha \beta }\,,
\label{TK}
\end{equation}%
or its inverse form
\begin{equation}
K_{\mu \alpha \beta }=\frac{1}{2}\,\left( T_{\alpha \mu \beta }-T_{\mu
\alpha \beta }+T_{\beta \mu \alpha }\right) \,.  \label{KT}
\end{equation}

We will examine a sector in the solution space, interesting for holography applied to condensed matter. Since we are looking for dislocations in crystalline systems, the simplest case that also has physical significance corresponds to axially symmetric and static $K^{ab}$. Holographic equations in $\omega ^{ab}$ (and therefore in $K^{ab}$) that we want to solve, given
by \eqref{constraints}, are complicated because of their non-linearity. To
simplify them, we need a physical ansatz on $\omega ^{ab}$, consistent with
the black hole metric.

The most interesting black holes for holographic purposes are black branes,
with flat horizons ($\varkappa =0$). To implement the axial
symmetry, we will write the three-dimensional transversal section of the
metric in cylindrical coordinates $y^{i}=(\rho ,\varphi,z)$, as
\begin{equation}
\de\Omega ^{2}=\de\rho ^{2}+\rho ^{2}\de\varphi ^{2}+\de z^{2}\,. \label{cyl}
\end{equation}
Planar black holes have isometries given by the six Killing vectors $\xi \in \{p_{m},j_{m}\}$, where $p_{m}$ are transversal translations and $j_{m}$ are transversal rotations, whose explicit expressions in cylindrical coordinates are given in Appendix \ref{conventions} (see Eq.~\eqref{Killing}).
They satisfy the algebra \eqref{Kalg} with non-vanishing Lie brackets
\begin{equation}
[ j_{m},j_{n}]=-\epsilon _{mnk}\,j_{k}\,,\qquad [j_{m},p_{n}]=\epsilon _{mnl}\,p_{k}\,.
\end{equation}
In addition, a static system has also the Killing vector $p_0\partial _t$ that commutes with all previous isometries.

To introduce a dislocation in the field theory, we break the static spherical symmetry of the metric into the static axial symmetry,
$\{p_{0},p_{m},j_{m}\}\rightarrow \{p_{0},j_{3}\}$ of the torsion field. The isometry $j_{3}=\partial _{\varphi }$ corresponds to the invariance under rotations around the $z$-axis. 

The contorsion tensor has isometries $\xi \in \{p_{0},j_{3}\}$ if it satisfies
\begin{equation}
\pounds _{\xi }K^{ab}=\left( \xi ^{\alpha }\partial _{\alpha }K_{\ \ \lambda
}^{ab}+\partial _{\lambda }\xi ^{\alpha }K_{\ \ \alpha }^{ab}\right) \mathrm{d}x^{\lambda }=0\,,
\end{equation}
yielding the solution
\begin{equation}
K^{ab}=K_{\ \ \mu }^{ab}(\rho ,z)\,\mathrm{d}x^{\mu }\,. \label{4cont}
\end{equation}
Similarly, for the torsion, the vanishing Lie derivative
\begin{equation}
\pounds _{\xi }T^{a}=\left( \xi ^{\alpha }\partial _{\alpha }T_{\ \mu \nu
}^{a}+\partial _{\mu }\xi ^{\alpha }T_{\ \ \alpha \nu }^{a}+\partial _{\nu
}\xi ^{\alpha }T_{\ \mu \alpha }^{a}\right) \mathrm{d}x^{\mu }\wedge \mathrm{d}x^{\nu }=0\,,
\end{equation}
implies $T^{a}=\frac{1}{2}\,T_{\mu \nu }^{a}(\rho ,z)\,\mathrm{d}x^{\mu
}\wedge \mathrm{d}x^{\nu }$, which is consistent with $K^{ab}$.

Finally, we recall that the general form of a 
 rank-three tensor in four
dimensions, with two antisymmetric indices, is given by a decomposition to its irreducible components
\begin{equation}
T_{\mu \nu \alpha }=-T_{\mu \alpha \nu }=\sqrt{|g|}\,\epsilon _{\mu \nu
\alpha \beta }\,A^{\beta }+B_{\alpha }g_{\mu \nu }-B_{\nu }g_{\mu \alpha
}+\tau _{\mu [ \nu \alpha ]}\,,
\end{equation}
where $g=\det[g_{\mu\nu}]$, $A^{\mu }$ is the \textit{axial torsion} vector describing the completely antisymmetric part of the torsion (4 components), 
\begin{equation}
A^{\mu }=\frac{1}{3!\sqrt{|g|}}\,\epsilon ^{\mu \nu \alpha \beta }T_{\nu
\alpha \beta }\,, \label{axial torsion}
\end{equation}
$B_{\mu }$ is the \textit{diagonal torsion} vector, corresponding to the trace of the torsion (4 components), 
\begin{equation}
B_{\mu }=\frac{1}{3}\,g^{\alpha \beta }T_{\alpha \beta \mu }\,,
\end{equation}%
while the \textit{tensorial torsion} $\tau _{\mu [ \nu \alpha ]}=-\tau _{\mu [\alpha \nu ]}$ is a traceless and non-axial tensor (16 independent components),
\begin{equation}
\epsilon ^{\alpha \beta \mu \nu }\tau _{\beta [ \mu \nu ]}=0\,,\qquad
g^{\mu \nu }\tau _{\mu [ \nu \alpha ]}=0\,.
\end{equation}
Similarly, this decomposition can be carried out for the contorsion field using 
\eqref{KT},
\begin{equation}
K_{\mu \nu \alpha }=-K_{\nu \mu \alpha }=-\frac{1}{2}\sqrt{|g|}\,\epsilon
_{\mu \nu \alpha \beta }\,A^{\beta }-B_{\mu }g_{\nu \alpha }+B_{\nu }g_{\mu
\alpha }+\frac{1}{2}\,\left( \tau _{\nu [ \mu \alpha ]}-\tau _{\mu
[ \nu \alpha ]}+\tau _{\alpha [ \mu \nu ]}\right) \,.
\end{equation}
We will restrict to the cases where $\tau _{\mu [ \nu \alpha ]}=0$.
Therefore, the axial torsion $A^{\mu }$ and the diagonal one $B_{\mu }$ are
the only nontrivial components of $T^{a}$ and $K^{ab}$,
\begin{eqnarray}
T_{\mu \nu \alpha } &=&\sqrt{|g|}\,\epsilon _{\mu \nu \alpha \beta
}\,A^{\beta }+B_{\alpha }g_{\mu \nu }-B_{\nu }g_{\mu \alpha }\,,  \notag \\
K_{\mu \nu \alpha } &=&-\frac{1}{2}\sqrt{|g|}\,\epsilon _{\mu \nu \alpha
\beta }\,A^{\beta }-B_{\mu }g_{\nu \alpha }+B_{\nu }g_{\mu \alpha }\,,
\label{A,B}
\end{eqnarray}
carrying all torsional degrees of freedom in the theory.

\subsection{Holographic Ward identities}
\label{Anomaly}

Observable quantities are introduced in a holographic QFT as $n$-point functions computed from the quantum effective action \eqref{eff}. In particular, holographic currents corresponding to the Poincar\'{e} symmetry of the theory are 1-point functions
\begin{equation}
\left\langle \tau _{~a}^{\mu }(x)\right\rangle =\frac{1}{\sqrt{\left\vert
g\right\vert }}\frac{\delta W[e,\omega ]}{\delta e_{\mu }^{a}(x)}\,,\qquad
\left\langle \sigma _{~ab}^{\mu }(x)\right\rangle =\frac{1}{\sqrt{\left\vert
g\right\vert }}\frac{\delta W[e,\omega ]}{\delta \omega _{\mu }^{ab}(x)}\,,
\end{equation}
where ${\tau ^{\mu }}_{a}$ is the energy-momentum tensor and $\sigma_{~ab}^{\mu}$ is the spin current. They are obtained from the variation 
\begin{equation}
\delta W=\int\limits_{\partial \mathcal{M}} \left(  \delta e^{a} \wedge\tau _{a}+\frac{1}{2}\,  \delta \omega ^{ab} \wedge  \sigma _{ab} \right)\,, \label{var_W}
\end{equation}
equivalent to \eqref{delta W}, only written in terms of differential forms, such that the 3-forms $\tau _{a}$ and $\sigma _{ab}$ are Hodge duals to the stress tensor and spin current, respectively. Explicitly, we have $\tau_{~a}^{\mu}=-\frac{1}{3!\sqrt{|g|}}\,\epsilon^{\mu\nu\alpha\beta}\tau_{a\,\nu\alpha\beta}$ and $\sigma_{~ab}^{\mu}=-\frac{1}{3!\sqrt{|g|}}\,\epsilon^{\mu\nu\alpha\beta}\sigma_{ab\,\nu\alpha\beta}$.
The variation \eqref{var_W} is computed from the renormalized CS AdS gravity action evaluated on-shell in Ref.~\cite{Banados:2006fe}. As a result,  the conserved currents are found in terms of the gravitational quantities as
\begin{eqnarray}
\left\langle \tau _{a}\right\rangle &=&-8\kappa\,\epsilon_{abcd}\left( R^{bc}+2\,e^{b}\wedge k^{c}\right) \wedge k^{d}\,,  \notag  \\
\left\langle \sigma _{ab}\right\rangle &=&-16\kappa \,\epsilon _{abcd}\,T^{c}\wedge k^{d}\,. \label{1-point}
\end{eqnarray}
They satisfy the quantum conservation laws \eqref{conservation} (Ward's identities),
associated with diffeomorphisms and Lorentz transformations. Furthermore, the energy-momentum tensor possesses the Weyl anomaly, proportional to the Euler invariant \cite{Henningson:1998gx,Schwimmer:2003eq,Schwimmer:2000cu,Deser:1993yx},
\begin{equation}
\left\langle \tau _{~a}^{a}\right\rangle =-\frac{\kappa }{4}\,\epsilon
^{\mu \nu \lambda \sigma }\,\epsilon _{\alpha \beta \gamma \delta }\,R_{\mu
\nu }^{\alpha \beta }\,R_{\lambda \sigma }^{\gamma \delta }\,,
\end{equation}
because the classical conservation law for the Weyl dilatations is $\tau_{~a}^{a}= e_\mu^a \tau_{~a}^{\mu}=0$.

We also want to explore the quantum properties of the current \eqref{J}, whose holographic form, using \eqref{1-point}, is given by
\begin{equation}
\left\langle J_{\mathrm{odd}}^{\mu }\right\rangle =\frac{8\kappa }{3\sqrt{|g|%
}}\,\epsilon ^{\lambda \nu \alpha \beta }\,\left( k_{\ \nu }^{\mu
}T_{\lambda \alpha \beta }-k_{\lambda \nu }T_{\ \alpha \beta }^{\mu }\right)
.
\end{equation}
In turn, the corresponding  holographic anomaly, expressed in terms of gravitational quantities, reads
\begin{equation}
\mathcal{A}_{\mathrm{odd}}=\frac{8\kappa }{3\sqrt{|g|}}\,\epsilon ^{\lambda
\nu \alpha \beta }\partial _{\mu }\left( k_{\ \nu }^{\mu }T_{\lambda \alpha
\beta }-k_{\lambda \nu }T_{\ \alpha \beta }^{\mu }\right) \,.  \label{Ach}
\end{equation}
 Locally, it is a total derivative, but if its global class is nontrivial, it describes a quantum anomaly. They are typically proportional to Chern characters. On a Riemannian manifold, a natural candidate is the Pontryagin density of the Lorentz group, $R^{ab}\wedge R_{ab}$. 
In a Riemann–Cartan space, however, the relevant candidates include the second Chern character of the AdS$_{4}$ group, $R^{ab}\wedge
R_{ab}+2\left( R^{ab}\wedge e_{a}\wedge e_{b}-T^{a}\wedge T_{a}\right) $,
and the Nieh-Yan invariant \cite{Nieh:1981ww}. There is an ongoing discussion in the
literature about whether these invariants contribute to the chiral
anomaly \cite{Chandia:1997hu,Aldazabal:1999nu,Chandia:1999az}.

In our case of the dimensionally continued black holes, the only torsion component relevant for the considered anomaly is the axial torsion $A^{\mu }$, such that, for the planar black hole with the transverse section \eqref{cyl}, it acquires the form
\begin{equation}
\mathcal{A}_{\mathrm{odd}}=-\frac{16\kappa M}{\rho }\,\partial _{i}\left(
\rho A^{i}\right) \,.  \label{anomaly}
\end{equation}
It is generated only by the axial part \eqref{axial torsion} of the spin 
connection. It remains to solve it from \eqref{constraints} and analyze its global properties.

\section{Purely axial holographic dislocation}
\label{Sec:Holographic-WSM}

So far, we have constructed the holographic QFT at the finite temperature applying the holographic dictionary to the dimensionally continued black hole. This setting completely determines the boundary 1-forms $e^a$ and $k^a$.
In our case with symmetric $k_{\mu \nu }$, such that $k^{a}\wedge e_{a}=0$, equations \eqref{constraints} simplify to
\begin{equation}
\begin{array}{llll}
C & =\epsilon _{abcd}\,F^{ab}\wedge F^{cd}\,, & C_{a} & =\epsilon
_{abcd}\,F^{bc}\wedge T^{d}\,,\medskip  \\
C_{ab} & =2\epsilon _{abcd}\,T^{c}\wedge \mathrm{D}k^{d}\,,\qquad \qquad & \bar{C}_{a} & =\epsilon _{abcd}\,F^{bc}\wedge \mathrm{D}k^{d}\,,
\end{array} \label{C}
\end{equation}
where
\begin{eqnarray}
\mathrm{D}k^{a} &=&\mathrm{\mathring{D}}k^{a}+K^{ab}\wedge k_{b}\,,  \notag
\\
F^{ab} &=&\mathrm{\mathring{D}}K^{ab}+K^{ac}\wedge
K_{c}^{\ b}+2\left( e^{a}\wedge k^{b}-e^{b}\wedge
k^{a}\right) \,.
\end{eqnarray}
In the above expressions, we use the fact that, in the planar case, Eq.~\eqref{omega} becomes
\begin{equation}
\mathring{\omega}^{ab}=\mathring{\omega}^{12} \,\delta^{ab}_{12}=-\delta^{ab}_{12}\,\mathrm{d}\varphi \quad \Rightarrow \quad \mathring{R}^{ab}=0\,. 
\end{equation}

Each equation in \eqref{C} is a 4-form, therefore proportional to the volume element \eqref{volume 4D}, becoming in that way a 0-form equation. The notation for the three-dimensional Levi-Civita symbol is given by  \eqref{Levi 3D}.
The holographic Eqs.~\eqref{C} determine the spin connection in terms of the axial torsion $A^\mu$ and the diagonal torsion $B^\mu$, as given by \eqref{A,B}.
The goal is to solve them in the unknown quantities  $A^\mu$ and $B^\mu$. 
This task is non-trivial because Eqs.~\eqref{C} are non-linear, as they have the origin in a non-linear gravitational theory.

As the first step, we write Eqs.~\eqref{C} in the tensorial form and, at the same time, decompose them in $x^\mu=(t,y^i)$, where $y^i=(\rho,\varphi,z)$ are cylindrical coordinates of the flat horizon. The Lorentz indices are decomposed in the tangent space as $a=(0,m)$, $m=1,2,3$. Because both tangent and spacetime indices use Latin characters, we will distribute them such that the curved indices start at the beginning of the alphabet, $i,j,k,l,\ldots$, while the flat indices start in the middle of the alphabet, $m,n,s,p,q,\ldots$.

Notice that the 3D flat space is without torsion, namely,
\begin{equation}    \tilde{T}^{m}=\mathrm{d}\tilde{e}^{m}+\mathring{\omega}^{mn}\wedge \tilde{e}_{n}=0\,.
\end{equation}
As a consequence, we have the identity
\begin{eqnarray}
\mathrm{\mathring{D}}k^{a} &=&\frac{M}{2}\, \delta^a_m \tilde T^m =0\,.  
\end{eqnarray}
All other 1-forms and 2-forms that we need to evaluate are given in Eqs.~\eqref{Kab}--\eqref{F} in Appendix \ref{Computations}. To reduce the four-dimensional Levi-Civita symbol to three dimensions, we use the convention $\epsilon _{tijk}\equiv \epsilon _{ijk}$, and denote the 3D surface element as $\mathrm{d}\sigma _{i}=\dfrac{\rho }{2}\,\epsilon _{ijk}\,\mathrm{d}y^{j}\wedge \mathrm{d}y^{k}$.
All notation is summarized in Appendix \ref{conventions}.\medskip

We have to analyze Eqs.~\eqref{C} in different branches of solutions. A simple possibility is to assume that the dislocation is purely axial, and try to solve the holographic equations when the diagonal torsion vector is zero, $B_{\mu }=0$. This case has been worked out in detail in Appendix \ref{Only A}. We found that the equations $\bar{C}_{a}=0$ lead to the branching of solutions.

When the axial torsion $A_{i}$ has only the horizontal component, $A_\rho$, and $A_z=0$ (see Appendix \ref{Horizontal}), there is a complex solution given by Eqs.~\eqref{sol}. This solution is not physical because it is complex, and also because it depends on an arbitrary function $\zeta (z)$, not determined uniquely by the holographic equations.

In the second branch, when $A_z \neq 0$ (see Appendix \ref{Nonhorizontal}), the solution is again complex and possesses two arbitrary functions. 

Therefore, a purely axial holographic dislocation does not exist. A holographic description of a dislocation requires a nontrivial diagonal torsion, $B_{\mu }\neq 0$. We focus on this case in the next section.

\section{Holographic dislocation with diagonal torsion} \label{WSM}

In this section, we find solutions with a non-trivial dislocation. First, we look at the equation $C_{ab}=0$ in \eqref{C}, with the  components
\begin{eqnarray}
C_{0m} &=&-\frac{M}{2}\,\epsilon _{mns}\,\mathrm{d}t\wedge \left(
\frac{1}{4}\,A^{t}\,\mathrm{d}\tilde{\sigma}^{n}-\tilde{e}^{n}\wedge
\tilde{B}\right) \wedge \left( \frac{1}{4}\,\tilde{A}^{s}+2B_{t}\,\tilde{e}^{s}\right) \,,  \notag \\
C_{mn} &=&-\frac{M}{4}\,\epsilon _{mns}\,B_{t}\,\mathrm{d}t\wedge \tilde{A}%
\wedge \tilde{e}^{s}\,,  
\end{eqnarray}
where $\tilde{B}\wedge \tilde{B}=0$ and $\tilde{A}\wedge \tilde{A}^{s}=0$ for symmetry reasons. See the end of Appendix~\ref{conventions} for a notation summary. The last equation can be solved when $A^{i}\neq 0$, leading to
\begin{equation}
 \quad B_{t}=0\,.
\end{equation}
The first equation, in this case, implies
\begin{eqnarray}
\frac{1}{2}\,A^{t}\mathcal{A}_{m}+\tilde{B}^{n}\tilde{A}_{nm}=0\,. \label{colineal}
\end{eqnarray}
A consistency relation is found by contracting it with $ \tilde{e}_i^m A^i$, yielding
\begin{equation}
 A^{t}=0\,. 
\end{equation}%
Then \eqref{colineal} implies that
\begin{equation}
 \epsilon ^{ijk}B_{j}A_{k}=0\,,  \label{C0m}
\end{equation}
namely, the 3D vectors $A_i=\gamma_{ij}A^j$ and $B_i$ satisfy $\vec{A} \times \vec{B}=0$, thus they are parallel. We will choose a particular solution where the proportionality factor between the vectors is constant,
\begin{equation}
  A_i =8c\,B_i\,,\qquad c=\mathrm{const.}\neq 0\,. \label{f(c)}
\end{equation}
We will call $c$ the \textit{dislocation parameter} because it describes the strength of the torsion field, as well as its internal structure, that is, how the diagonal torsion vector is twisted with respect to the axial torsion vector. Note that the equation \eqref{f(c)} allows some components of $A_i$ and $B_i$ to vanish, but not all of them, because $\gamma_{ij} A^i A^j \neq 0$.

In that case, the equation $C_{0}=0$ is identically satisfied, while the remaining ones become
\begin{eqnarray}
\bar{C}_{0} &\propto &\epsilon ^{ijk}B_{i}\partial _{j}B_{k}\,,  \notag \\
\bar{C}_{i} &\propto &\partial _{i}B^{2}-2\left[ \mathrm{\mathring{D}}_{n}
\tilde{B}^{n}-M+3\left( 1-c^{2}\right) B^{2}\right] B_{i}-4c^{2}\tilde{B}^{j}
\tilde{e}_{i}^{m}\mathrm{\mathring{D}}_{j}\tilde{B}_{m}\,,  \notag \\
C_{i} &\propto &\partial _{i}B^{2}-\left[ 2\mathrm{\mathring{D}}_{n}\tilde{B}^{n}-M+3\left( 1-c^{2}\right) \,B^{2}\right] B_{i}-2c^{2}\tilde{e}_{i}^{m}
\mathrm{\mathring{D}}_{n}\left( \tilde{B}_{m}\tilde{B}^{n}\right) \,,  \notag
\\
C &\propto &\left( c^{2}-1\right) \left[ \left( \mathrm{\mathring{D}}_{m}
\tilde{B}^{m}\right) ^{2}+\mathrm{\mathring{D}}_{n}\tilde{B}^{m}\mathrm{\mathring{D}}_{m}\tilde{B}^{n}-2MB^{2}\right] \label{Eqs.c} \\
&&+\left( 3c^{2}-1\right) \left( \tilde{B}^{i}\partial _{i}B^{2}+B^{2}
\mathrm{\mathring{D}}_{m}\tilde{B}^{m}\right) +M\mathrm{\mathring{D}}_{m}
\tilde{B}^{m}\,,  \notag
\end{eqnarray}
where we defined $B^2=\gamma^{ij} B_iB_j$. 
The form of the above equations simplifies when $c=\pm 1$, so we will start with the analysis of these cases first.

\subsection{`No-go' for the dislocation \texorpdfstring{$c^{2}=1$}{c2=1}}
\label{c2=1}

When $c^2=1$, Eqs.~\eqref{Eqs.c} are simplified to
\begin{eqnarray}
0 &=&\partial _{i}B^{2}-2\left( \mathrm{\mathring{D}}_{n}\tilde{B}^{n}-M\right) B_{i}-4\tilde{B}^{j}\tilde{e}_{i}^{m}\mathrm{\mathring{D}}_{j}%
\tilde{B}_{m}\,,  \notag \\
0 &=&\partial _{i}B^{2}-\left( 2\mathrm{\mathring{D}}_{n}\tilde{B}%
^{n}-M\right) B_{i}-2\tilde{e}_{i}^{m}\mathrm{\mathring{D}}_{n}\left( \tilde{%
B}_{m}\tilde{B}^{n}\right) \,, \\
0 &=&2\tilde{B}^{i}\partial _{i}B^{2}+2B^{2}\mathrm{\mathring{D}}_{m}\tilde{B%
}^{m}+M\mathrm{\mathring{D}}_{m}\tilde{B}^{m}\,.  \notag
\end{eqnarray}
To check their consistency, we contract the first two equations with $\tilde{B}^{i}$, yielding
\begin{eqnarray}
0 &=&\tilde{B}^{i}\partial _{i}B^{2}+2\left( \mathrm{\mathring{D}}_{n}\tilde{%
B}^{n}-M\right) B^{2}\,,  \notag \\
0 &=&\left( 4\mathrm{\mathring{D}}_{n}\tilde{B}^{n}-M\right) B^{2}\,, \\
0 &=&2\tilde{B}^{i}\partial _{i}B^{2}+\left( 2B^{2}+M\right) \mathrm{\mathring{D}}_{n}\tilde{B}^{n}\,.  \notag
\end{eqnarray}
Since $B^{2}\neq 0$, the second equation leads to $\mathrm{\mathring{D}}_{n} \tilde{B}^{n}=\frac{M}{4}$. Because the black hole mass parameter must satisfy the strict inequality $M> 0$ for a QFT at finite temperature, as seen from \eqref{T}, the other two equations become
\begin{equation}
\tilde{B}^{i}\partial _{i}B^{2}=\frac{3M}{2}\,B^{2}\,,\qquad B^{2}=-\frac{M}{14}\,.
\end{equation}
This result is non-physical because it corresponds to the bulk geometry that is a naked singularity ($M<0$) at the zero temperature. Furthermore, constant value of $B^{2}$ leads to the inconsistent equation $\frac{3M}{2}B^{2}= 0$. 

The only possible solution in the torsion field is the trivial one. Since we want to describe dislocations, we will assume $c^2\neq 1$ in the rest of the text.

\subsection{Solution with a generic dislocation \texorpdfstring{$c^{2}\neq 1$}{c2=1}}
\label{c2n1}

When the dislocation parameter satisfies $c^{2}\neq 0$ and $c^{2}\neq 1$, we
can treat first two equations of \eqref{Eqs.c} as algebraic equations in $%
B_{i}\mathrm{\mathring{D}}_{n}\tilde{B}^{n}$ and $\tilde{e}_{i}^{m}\tilde{B}%
^{j}\mathrm{\mathring{D}}_{j}\tilde{B}_{m}$, and solve them as
\begin{equation}
B_{i}\mathrm{\mathring{D}}_{n}\tilde{B}^{n}=a\partial _{i}B^{2}\,,\qquad
\tilde{e}_{i}^{m}\tilde{B}^{j}\mathrm{\mathring{D}}_{j}\tilde{B}_{m}=a\,\partial _{i}B^{2}+( m+bB^{2}) \,B_{i}\,,
\label{XY}
\end{equation}
where we introduced the constants
\begin{equation}
m=\frac{M}{2c^{2}}>0\,,\qquad a=\frac{1}{2\left( 1+2c^{2}\right) }>0
\,,\qquad b=\frac{3\left( c^{2}-1\right) }{2c^{2}}\neq 0\,. \label{a,b}
\end{equation}
From the contractions  of Eqs.~\eqref{XY} with $\tilde{B}^{i}$,
we find the useful identities mapping the differential expressions to the algebraic ones,
\begin{equation}
\mathrm{\mathring{D}}_{n}\tilde{B}^{n}=\frac{m+bB^{2}}{2c^{2}}
\,,\qquad \tilde{B}^{i}\partial _{i}B^{2}=\frac{(
m+bB^{2}) B^{2}}{2ac^{2}}\,,  \label{XY1}
\end{equation}
where we applied the identity $1-2a=4ac^{2}$. Consequently,
\begin{equation}
\partial _{i}B^{2}=\frac{m+bB^{2}}{2ac^{2}}\,B_{i}\,,\qquad \tilde{e}%
_{i}^{m}\tilde{B}^{j}\mathrm{\mathring{D}}_{j}\tilde{B}_{m}=\frac{%
m+bB^{2}}{4ac^{2}}\,B_{i}\,.  \label{XY2}
\end{equation}%
Note that (\ref{XY2}) is just a consequence of the previous equations, and not
equivalent to them. Thus, when (\ref{XY2}) is satisfied, we still have to
check the first equation of (\ref{XY1}), while the second one is
automatically satisfied.

The last equation in \eqref{Eqs.c} yields
\begin{equation}
\mathrm{\mathring{D}}_{m}\tilde{B}^{n}\mathrm{\mathring{D}}_{n}\tilde{B}%
^{m}=\alpha B^{4}+\beta B^{2}+\gamma \,,  \label{DBDB}
\end{equation}%
with the coefficients
\begin{eqnarray}
\alpha &=&\frac{3}{4c^{4}}\left[ \frac{3\left( c^{2}-1\right) ^{2}}{4c^{4}}%
-\left( 1-3c^{2}\right) \left( 3+4c^{2}\right) \right] \,,  \notag \\
\beta &=&M\left[ \frac{3\left( c^{2}-1\right) }{8c^{8}}-\frac{\left(
1-3c^{2}\right) \left( 3+4c^{2}\right) }{4c^{4}\left( c^{2}-1\right) }+\frac{%
3}{4c^{4}}-2\right] \,,  \label{alpha,beta,gamma} \\
\gamma &=&\frac{M^{2}}{4c^{4}}\left( \frac{1}{4c^{4}}-\frac{1}{1-c^{2}}
\right) \,.
\notag
\end{eqnarray}

We have to solve these equations in the diagonal torsion $B_{i}(\rho ,z)$.
In components, for the first equation (\ref{XY2}), we find%
\begin{equation}
B_{\varphi }=0\,,\qquad \partial _{\rho }B^{2}=\frac{m+bB^{2}}{%
2ac^{2}}\,B_{\rho }\,,\qquad \partial _{z}B^{2}=\frac{m+bB^{2}}{%
2ac^{2}}\,B_{z}\,,  \label{Xi}
\end{equation}%
where $B^{2}=B_{\rho }^{2}+B_{z}^{2}$. 
Thanks to $B_{\varphi }=0$ and $\partial _{\varphi }=0$, the equation $%
\epsilon ^{ijk}B_{i}\partial _{j}B_{k}=0$ is identically satisfied, while
the second equation (\ref{XY2}) reads in components%
\begin{eqnarray}
B_{\rho }\partial _{\rho }B_{\rho }+B_{z}\partial _{z}B_{\rho } &=&\frac{%
m+bB^{2}}{4ac^{2}}\,B_{\rho }\,,  \notag \\
B_{\rho }\partial _{\rho }B_{z}+B_{z}\partial _{z}B_{z} &=&\frac{%
m+bB^{2}}{4ac^{2}}\,B_{z}\,.  \label{Yi}
\end{eqnarray}
Taking the difference of one (\ref{Xi}) and two (\ref{Yi}), we obtain a simpler system
\begin{eqnarray}
B_{z}\left( \partial _{\rho }B_{z}-\partial _{z}B_{\rho }\right) &=&0\,,
\notag \\
B_{\rho }\left( \partial _{\rho }B_{z}-\partial _{z}B_{\rho }\right) &=&0\,,
\end{eqnarray}
where each equation becomes factorized. Thus, the result depends on which factor vanishes.  

When one component of $B_i$ is vanishing, say $B_{z}=0$, $B_{\rho }=B_{\rho }(\rho )$, the first differential equation in \eqref{XY} reduces to $(1-2a)B_{\rho }\frac{\de B_{\rho }}{\de\rho }=0$, which is consistent only when $2a=1$,$\,$or equivalently $c=0$. Since $c\neq
0$, this case is not allowed.

The case with another component vanishing, $B_{\rho }=0$, $B_{z}=B_{z}(z)$,
is equivalent, with the replacement $\rho \rightarrow z$, such that there is
no solution in this case either.

Therefore, the only allowed possibility corresponds to both components
non-vanishing, $B_{z}B_{\rho }\neq 0$. Then, it must hold
\begin{equation}
\partial _{\rho }B_{z}=\partial _{z}B_{\rho }\qquad \Leftrightarrow \qquad
\nabla \times \vec{B}=0\,,  \label{dBz=dBr}
\end{equation}%
meaning that the field $\vec{B}$ is irrotational and therefore it is a gradient of some torsion potential,
\begin{equation}
\vec{B}=\nabla \psi \qquad \Leftrightarrow \qquad B_{i}=\partial _{i}\psi \,.
\label{grad}
\end{equation}%
This leaves only two independent equations,%
\begin{eqnarray}
B_{\rho }\partial _{\rho }B_{\rho }+B_{z}\partial _{z}B_{\rho } &=&\frac{%
m+b\left( B_{\rho }^{2}+B_{z}^{2}\right) }{4ac^{2}}\,B_{\rho }\,,
\notag \\
B_{\rho }\partial _{z}B_{\rho }+B_{z}\partial _{z}B_{z} &=&\frac{%
m+b\left( B_{\rho }^{2}+B_{z}^{2}\right) }{4ac^{2}}\,B_{z}\,.
\label{rama3}
\end{eqnarray}

Finally, from \eqref{Xi}, it is clear that there are two branches of solutions,
corresponding to either a constant norm of the diagonal torsion, $B^{2}=-\frac{m}{b}$, or an arbitrary norm of the diagonal torsion, $B^{2}\neq const$.
It turns out that only one of them permits a non-trivial solution, as we show next.

\subsubsection{`No-go' for the constant norm \texorpdfstring{$B^2 = const$}{B2=const}}

Assuming that the vector  $B_i$ has a constant norm, its value is
\begin{equation}
|B|\equiv\sqrt{B^{2}_{\rho} + B^{2}_{z}} =\sqrt{-\frac{m}{b}}\,,\qquad 1-c^{2}>0\,,
\end{equation}
and a general solution for the components reads 
\begin{equation}
B_\rho=\sqrt{-\frac{m}{b}}\,\sin \vartheta \,,\qquad B_z=
\sqrt{-\frac{m}{b}}\,\cos \vartheta \,,\qquad \vartheta =\vartheta (\rho ,z)\,. \label{vartheta}
\end{equation}
From \eqref{XY1}, we can deduce  the identities   $\mathrm{\mathring{D}}_{n} \tilde{B}^{n}=0$ and $\tilde{B}^{i}\partial _{i}B^{2}=0$, which allow us to write the last equation in \eqref{Eqs.c} as
\begin{equation}
\mathrm{\mathring{D}}_{m}\tilde{B}^{n}\mathrm{\mathring{D}}_{n}\tilde{B}^{m}=-2 MB^{2}\,. \label{1}
\end{equation}
The l.h.s.~of the above expression is computed  directly from the definition of the covariant derivative,
\begin{equation}
\mathrm{\mathring{D}}_{m}\tilde{B}^{n}\mathrm{\mathring{D}}_{n}\tilde{B}^{m}=\left( \partial _{\rho }\tilde{B}^{1}\right) ^{2}+\left( \partial _{z}\tilde{B}^{3}\right) ^{2}+2\partial _{\rho }\tilde{B}^{3}\partial _{z}\tilde{B}^{1}+\left( \frac{1}{\rho }\,\tilde{B}^{1}\right) ^{2}\,.  \label{DBDB2}
\end{equation}
Plugging it back into \eqref{1} and using the solution \eqref{vartheta}, we obtain
\begin{equation}
\cos ^{2}\vartheta \left( \partial _{\rho }\vartheta \right) ^{2}+\sin ^{2}\vartheta \left( \left( \partial _{z}\vartheta \right) ^{2}+\frac{1}{\rho ^{2}}\right)-2\sin \vartheta \cos \vartheta \,\partial _{\rho }\vartheta \partial_{z}\vartheta =-2M\,. \label{DBDBfin}
\end{equation}
Any solution for $\vartheta$, if it exists, is a solution of the complete system.
However, the vanishing covariant derivative identity enables us to find 
\begin{equation}
\mathring{D}_n\tilde{B}^n=0\quad \Rightarrow \quad \partial
_{\rho }\vartheta =\left( \partial _{z}\vartheta -\frac{1}{\rho }\right) \tan \vartheta \,.
\end{equation}
When $\sin\vartheta \neq 0$, substituting the expression for $\partial_z\vartheta$ into \eqref{DBDBfin}, all terms with $\partial_z\vartheta$ cancel out, such that
$\sin^{2}\vartheta = - 2 M \rho^2$.
This solution is not physical for real $\vartheta$ and positive $M$. 
The only possibility is thus to have  $\sin\vartheta = 0$, meaning $\vartheta=n\pi$ ($n\in \mathbb{Z}$), which implies $\tilde{B}^{1}=0$ and $\tilde{B}^{3}=const$, and the expression for \eqref{DBDB2} becomes
$\mathrm{\mathring{D}}_{m}\tilde{B}^{n}\mathrm{\mathring{D}}_{n}\tilde{B}^{m}=\left( \partial _{z}\tilde{B}^{3}\right) ^{2}$,
which results in an inconsistent expression for \eqref{DBDBfin}, namely $0=M$.

Therefore, it is concluded that the norm of the vector $B_i$ cannot be constant.

\subsubsection{Irrotational solution} \label{IRR}

Consider finally an irrotational vector $B_i = \partial_{i}\psi$ whose norm is not constant.
We have to solve Eqs.~\eqref{XY1}--\eqref{rama3}, where the second one is written in components as \eqref{rama3}.

The general solution is
\begin{equation}
\psi(\rho,z) =\psi _{0}-\frac{4ac^{2}}{b}\,\ln \sin \left( \frac{\sqrt{2mb}}{8ac^{2}}\,\left(
\rho +z\right) +\theta \right)\,,
\end{equation}
where  $\psi _{0}$ and $\theta $ are integration constants. We will set $\theta =0$ because the center of the coordinate system can always be shifted
along $z$ so that this is fulfilled. We will also set $\psi _{0}=0$ because
the torsional field is the derivative of $\psi $, thus this constant will not
contribute.

Because the root $\sqrt{2mb}$ can become complex for certain values of $c$, in terms of real functions, the general solution is
\begin{equation}
\psi =\left\{
\begin{array}{ll}
2\zeta \ln \sin \left( \rule[2pt]{0pt}{10pt}\omega \left( \rho
+z\right)  \right)  \,,\medskip  & c^{2}-1>0\,, \\
2\zeta \ln  \sinh\left( \rule[2pt]{0pt}{10pt}\omega \left(
\rho +z\right)  \right) \,, & c^{2}-1<0\,,
\end{array}
\right.
\end{equation}
where we introduced the parameters
\begin{equation}
\zeta =\frac{2c^{4}}{3\left( 1+2c^{2}\right) \left( 1-c^{2}\right) }\,,\qquad
\omega =\frac{1+2c^{2}}{4c^{4}}\sqrt{\frac{3M}{2}\,\left\vert
c^{2}-1\right\vert }>0\,.  \label{xi-omega}
\end{equation}
The potential  $\psi \left( \rho,z \right)$ depends only on one
variable, $\rho +z$. The symmetry $\rho \leftrightarrow z$ implies the equality of the components,
\begin{equation}
\label{IrrB}
B_{\rho }=B_{z}=\left\{
\begin{array}{ll}
2\omega \zeta \,\cot \left( \rule[2pt]{0pt}{10pt} \omega \left( \rho +z\right) \right)\,,\medskip  & c^{2}-1>0\,,  \\
2\omega \zeta \,\coth \left( \rule[2pt]{0pt}{10pt}\omega \left( \rho +z\right) \right) \,, &
c^{2}-1<0\,. 
\end{array}
\right.
\end{equation}
These two solutions are physically different because they have different periodicity, for instance.

We still have to check Eqs.~\eqref{XY1} and \eqref{DBDB}. Considering that the diagonal torsion  depends only on  $z+\rho$, we can rewrite them as 
\begin{eqnarray}
     2\partial_{\rho}B_{\rho} + \frac{1}{\rho^2}\,B_{\rho} &=& \, \frac{m+2bB_{\rho}^{2}}{2c^2}\,, \notag\\
4\left(\partial_{\rho}B_{\rho}\right)^2 + \frac{1}{\rho^4}\,B_{\rho}^{2} &=& 4\alpha B_{\rho}^{4} + 2\beta B_{\rho}^{2} + \gamma\,.  
\end{eqnarray}
A nicer form of these equations where a square of the derivative has been eliminated is
\begin{eqnarray}
2\partial _{\rho }B_{\rho } &=&-\frac{1}{\rho ^{2}}\,B_{\rho }+\frac{m+2bB_{\rho }^{2}}{2c^{2}}\,,  \notag \\
0 &=&\left( 4\alpha -\frac{b^{2}}{c^{4}}\right) B_{\rho }^{4}+\frac{2bB_{\rho }^{3}}{\rho ^{2}c^{2}}+\left( 2\beta -\frac{2}{\rho ^{4}}-
\frac{2mb}{c^{2}}\right) \,B_{\rho }^{2}+\frac{mB_{\rho }}{\rho
^{2}c^{2}}+\gamma -\frac{m^{2}}{4c^{4}}. \label{dB}
\end{eqnarray}

In the above equations, the terms $\frac{1}{\rho^{2}}$ break the symmetry between $\rho$ and $z$, so these equations cannot be fulfilled for all $(\rho,z)$.
Hence, we assume that there is a solution only at the ring $\mathcal{R}$ of the radius $\bar{\rho}$ located in the horizontal plane with the center at $\bar{z} = 0$, for which the radial component $B_{\rho }(\bar{\rho},\bar{z})=\Omega $ is a constant parameter. Then it is straightforward to show that $\Omega \neq 0$, otherwise the equations become inconsistent. Furthermore, taking the radial derivative of \eqref{IrrB} and evaluating it at the ring $\mathcal{R}$,  we can express $\partial _{\rho }B_{\rho }$ in terms of $\Omega $, finding
\begin{equation}
\text{At } \mathcal{R}:\quad \rho =\bar{\rho}\,,\quad z=0\,,\quad  B_{\rho }=\Omega \,,\quad  \partial _{\rho }B_{\rho}=-2\zeta \omega ^{2} \mathrm{sgn}(c^{2}-1)-\,\dfrac{\Omega ^{2}}{2\zeta }\,.  \label{Omega}
\end{equation}

With this at hand, the equations become algebraic,
\begin{eqnarray}
0 &=&\frac{3(c^2-1)}{c^2}\,\Omega^2 + \frac{\Omega}{\bar{\rho}^2} + m\,, \notag  \\
0 &=&\frac{3\left( 3c^{2}-1\right) \left( 4c^{2}+3\right) }{c^{4}}\,\Omega
^{4}+\frac{3\left( c^{2}-1\right) }{\bar{\rho}^{2}c^{4}}\,\Omega ^{3}+\frac{m\Omega }{\bar{\rho}^{2}c^{2}}-\frac{m^{2}}{1-c^{2}}   \label{poly} \\
&&+\left( \frac{3\left( c^{2}-1\right) \left( 1-2c^{2}\right) m}{2c^{6}}-\frac{\left( 1-3c^{2}\right) \left( 3+4c^{2}\right)m}{c^{2}\left(
c^{2}-1\right) }+\frac{3m}{c^{2}}-8c^{2}m-\frac{2}{\bar{\rho}^{4}}
\right) \,\Omega ^{2}\,,  \notag 
\end{eqnarray}
where we expressed all the quantities in terms of the four free parameters $(c,m,\Omega ,\bar{\rho})$, where $c$ is the dislocation parameter, $\Omega$ is the strength of the torsion field, $\bar{\rho}$ is the radius of the ring and  $m$ is the mass parameter that fixes the temperature of the holographic QFT as 
\begin{equation}
    m=\frac{2\pi^2}{c^2}\,T^2>0\,. \label{m-T}
\end{equation}

These parameters have to be solved such that \eqref{poly} is fulfilled.
The first equation leads to the following solution for the ring radius,
\begin{equation}
\frac{1}{\bar{\rho}^{2}}=\frac{3(1-c^2)}{c^2}\,\Omega -\frac{m}{\Omega }\,. 
\label{rho}
\end{equation}
When $c^{2}>1$, the r.h.s.~of the above equation is always positive when $\Omega $ is negative while, when $c^{2}<1$, there are two cases when the r.h.s.~becomes positive. This can be summarized by
\begin{equation}
\begin{array}{lll}
\bar{\rho}^{2}>0 & \quad \Rightarrow \quad  & 
\begin{array}{lll}
\Omega <0 \,, & m>0\,,\medskip  & c^{2}-1>0\,, \\
\Omega >0\,, \quad & 0<m<\frac{3(1-c^{2})}{c^{2}}\,\Omega ^{2} \quad\,,\medskip  &
c^{2}-1<0\,, \\
\Omega <0\,, & m>\frac{3(1-c^{2})}{c^{2}} \,\Omega ^{2}\,, & c^{2}-1<0\,,
\end{array}
\end{array}
\label{rho'}
\end{equation}
such that a positive solution $\bar{\rho}=\bar{\rho}(m,c,\Omega )$ always exists, according to the above inequalities. 
Replacing obtained $\frac{1}{\bar{\rho}^{2}}$ in the second equation \eqref{poly}, we can determine $\Omega $ from the polynomial%
\begin{equation}
P(\Omega )=P_{4}\,\Omega ^{4}+mP_{2}\,\Omega ^{2}+m^2 P_{0}=0\,,  \label{P}
\end{equation}
with the coefficients that depend only on $c$,
\begin{eqnarray}
P_{4} &=&\frac{3(6c^{6}+14c^{4}-3c^{2}-3)}{c^{6}}\,, \notag \\
P_{2} &=&-\frac{16c^{10}-16c^{8}-46c^{6}-3c^{4}+24c^{2}-3}{2c^{6}(c^{2}-1)}%
\,, \\
P_{0} &=&-\frac{2c^{4}-2c^{2}-1}{c^{2}(c^{2}-1)}\,.  \notag
\end{eqnarray}

The existence of real solutions of the above quadratic polynomial in $\Omega ^{2}$ depends on its discriminant 
$m^{2}\Delta= (mP_2)^2-4m^2 P_4  P_0 $, 
such that the normalized discriminant $\Delta $ is a function of the dislocation parameter only,
\begin{eqnarray}
\Delta  &=&\frac{1}{c^{12}(c^{2}-1)^{2}}\left( \rule[2pt]{0pt}{12pt}%
64c^{20}-128c^{18}-160c^{16}+392c^{14}+73c^{12}\right.   \notag \\
&&\left. +165c^{10}-\frac{999}{4}\,c^{8}-39c^{6}+\frac{225}{2}%
\,c^{4}-36c^{2}+\frac{9}{4}\right) \,.  \label{Delta}
\end{eqnarray}
Solving the quadratic equation \eqref{P} in $\Omega^2$, we obtain two solutions  linear in $m$,
\begin{equation}
\Omega ^{2}= m\,\frac{-P_2\pm \sqrt{\Delta}}{2P_4} \equiv m\,\Omega_{0,\pm}^{2}(c)\,, \label{Oo}
\end{equation}
where $\Omega_{0,\pm}$ is $m$-independent part. Its explicit form is
\begin{equation}
\Omega _{0,\pm }^{2}=\frac{16c^{10}-16c^{8}-46c^{6}-3c^{4}+24c^{2}-3\pm
2c^{6}|c^{2}-1|\sqrt{\Delta }}{12\left( c^{2}-1\right) \left(
6c^{6}+14c^{4}-3c^{2}-3\right) }\,.  \label{Opm}
\end{equation}
Then, the radius \eqref{rho} can be solved as
\begin{equation}
\frac{1}{\bar{\rho}^{2}}=\frac{\sqrt{m}}{R^{2}(c)}\quad \Rightarrow \quad \bar{\rho}=m^{-\frac{1}{4}}R(c)\,, \label{bR}
\end{equation}
with the $c$-dependent radial function given by
\begin{equation}
R(c)=\left( \frac{3\left(1- c^{2}\right) }{c^{2}}\,\Omega _{0,\pm }-\frac{1}{\Omega _{0,\pm }}\right) ^{-\frac{1}{2}}\,.  \label{R}
\end{equation}

The above expressions have dependence on $m$ in the torsion strength $\Omega$ completely determined, because the parameter $\omega$ can also be factorized as
\begin{equation}
    \omega = \sqrt{m}\,\omega_0(c)\,,\qquad \omega_0(c)=\frac{1+2c^{2}}{4c^{4}}\sqrt{3c^{2} | c^2 -1|}\,. \label{Om_0}
\end{equation}

To ensure the existence of physical solutions, the following conditions have to be fulfilled:
\begin{itemize}

\item[1)] $\Omega ^{2}$ \textit{is real}. This is satisfied when the discriminant $\Delta $, given by \eqref{Delta}, is non-negative, $\Delta \geq 0$.  This condition is fulfilled for all dislocations in the intervals $0<|c|\leq c_{01}\approx 0.288$ and $|c|\geq c_{02}\approx 0.685$, except in the points $|c|=1$, where it becomes divergent.

\item[2)] $\Omega $ \textit{is real}. This is analyzed from the positivity of $\Omega _{0,\pm }^{2}$ given by \eqref{Opm}.  This condition  is satisfied for $\Omega _{0,+}^{2}$ when the dislocations are in the intervals $c_{02}\leq |c|<c_{\infty} $ and $|c|>c_*$, where $c_{\infty }\approx
0.719$ is a divergence point due to the zero of the polynomial in the denominator, while $c_* \approx 1.169$ is its zero. On the other hand, $\Omega _{0,-}^{2}$ is
positive for the dislocations $c_{02}\leq |c|<1$, avoiding the divergent points at $|c|=1$.

\item[3)] $\bar{\rho}$ \textit{is real}. This condition has already been discussed from the positivity of $\bar{\rho}^{2}$, as given by Eqs.~\eqref{rho}. This requirement is satisfied under the conditions \eqref{rho'}.

\item[4)] \textit{Dependence on $m$ is consistent}. We determined the $m$-dependence in all the quantities previously discussed. The consistency of it is ensured by
comparison of the $m$-dependence in the l.h.s.~and r.h.s.~of the equation \eqref{Omega}. It turns out that these dependencies differ in general unless the mass itself is a function of the dislocation. This will give a particular relation between the temperature $T=\sqrt{\frac{m}{2}}\,\frac{|c|}{\pi }$  and  the dislocation parameter $c$. Finally, it should be checked whether obtained $m(c)$ (or $T(c)$)  satisfies the inequalities given by \eqref{rho'}.
\end{itemize}

We summarize the critical points $c$ of the solutions in the following table:
\begin{equation}
\fbox{$
\begin{array}{llllll}
c\medskip  & =0 &  &  & \rightarrow  & \text{excluded point;} \\
|c|\medskip  & =c_{01} & \approx  & 0.288 & \rightarrow  & \text{vanishing
discriminant, }\Delta =0\text{;} \\
|c|\medskip  & =c_{02} & \approx  & 0.685 & \rightarrow  & \text{vanishing
discriminant, }\Delta =0\text{;} \\
|c|\medskip  & =c_{\infty } & \approx  & 0.719 & \rightarrow  & \text{%
divergence of }\Omega _{0,+}^{2}\text{;} \\
|c|\medskip  & =1 &  &  & \rightarrow  & \text{excluded point;} \\
|c| & =c_{\ast } & \approx  & 1.169 & \rightarrow  & \text{zero of }\Omega
_{0,+}^{2}\text{.}
\end{array}
$} 
\label{tab:critical points}
\end{equation}

To ensure the fulfillment of the four conditions, we analyze first the periodic solutions, corresponding to the large values of the dislocation, $c^2-1>0$, and then the non-periodic ones, with the small values of the dislocation, $c^2-1<0$.

\paragraph{Periodic solution.}

Consider the case of the large dislocation parameter. We have already proved that the first three conditions are satisfied only in the positive branch of the torsion field when $\Omega_{0,+}<0$, for the dislocation parameter  $|c|>c_*$. 
The torsion dependence in the radial coordinate is given by \eqref{xi-omega}
and \eqref{IrrB} as
\begin{equation}
\Omega _{0,+}(c)=-\sqrt{\frac{c^{2}}{3(c^{2}-1) }}\cot \left(m^{\frac{1}{4}} \,\omega_0(c)R(c) \right)<0 \,,
\label{eq:periodic-sol}
\end{equation}
where we  replaced the known expression for $\bar \rho$ taken from \eqref{R}. The diagonal torsion field component is negative when the argument of the cotangent lies in the interval $\left(0,\frac{\pi}{2}\right)$. We restrict to the first period to have an invertible expression.

We still have to enforce the fourth condition. Because the function $\Omega _{0,+}(c)$ can depend explicitly only on the dislocation parameter, and not on the temperature, which scales as $\sqrt{m}$, the consistency of the above identity implies that $m$ must be a function of $c$.  Therefore, mass is not an independent parameter. Inverting the relations \eqref{eq:periodic-sol}, 
we obtain that the mass of the black hole, or the temperature $T=\frac{|c|}{\pi }\,\sqrt{\frac{m}{2}}$ of the QFT, is not arbitrary, but it depends on the
dislocation parameter  $|c|>c_*\approx 1.169$,
\begin{equation}
m=\frac{\mathrm{arccot}^4\left( - \sqrt{\frac{3\left(
c^{2}-1\right) }{c^{2}}}\,\Omega _{0,+}\right)}{\omega_0^4R^4}\,. 
\label{eq:mass-periodic}
\end{equation}
It can be checked straightforwardly that the argument of the cotangent lies in the required interval for any $c>c_*$.


\begin{figure}[t!]
    \centering
\includegraphics[width=\linewidth]{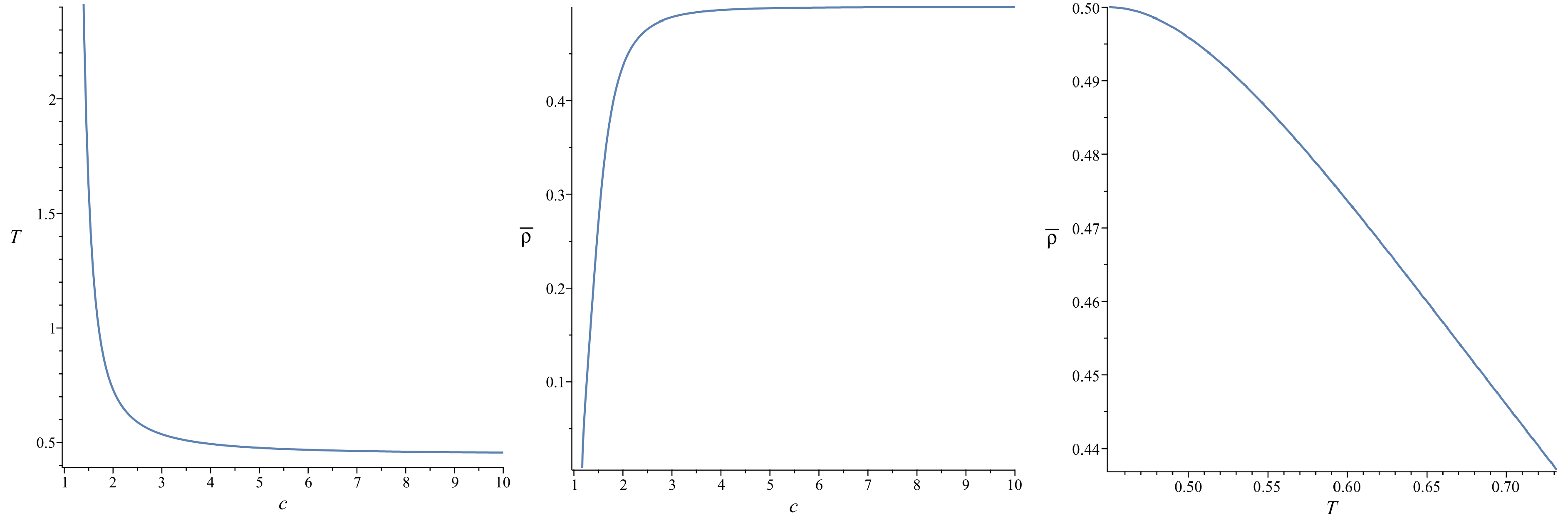}
    \caption{Periodic solution with the strength of the torsion field given by Eq.~\eqref{eq:periodic-sol}: (Left) Temperature, $T=\frac{|c|}{\pi}\sqrt{\frac{m}{2}}$, with the mass of the black hole in Eq.~\eqref{eq:mass-periodic}, as a function of the dislocation parameter, $c$. (Center) The radius of the ring,  $\bar\rho$, as given by Eqs.~\eqref{bR} and \eqref{R},  as a function of the dislocation parameter, $c$. (Right) Dependence of the radius from the temperature, obtained from integrating out the dislocation parameter numerically.}
    \label{Fig.periodic}
\end{figure}


To understand the physical behavior of the system, the temperature $T$ and the radius of the ring $\bar{\rho}=m^{-\frac{1}{4}}R$ of the crystalline material, as functions of the dislocation parameter, as well as $\bar\rho(T)$, are shown in Fig.~\ref{Fig.periodic} in the interval\footnote{The interval $c<-c_*$ is obtained from the parity of $T(c)$ and $\bar{\rho}(c)$.} $c>c_*$. Both $T$ and $\bar{\rho}$ are monotonous functions of $c$. While the ring radius, shown in Fig.~\ref{Fig.periodic} (center), increases, the temperature shown in Fig.~\ref{Fig.periodic} (left) decreases, as the torsion (dislocation parameter) increases.
For weak torsion, close to $c_*$, the temperature becomes very high, making the ring very small. This can also be concluded from Fig.~\ref{Fig.periodic} (right) displaying the dependence of the ring radius on the temperature,  confirming that larger rings correspond to lower temperatures. In turn, in
the strong-twisting limit, $c \rightarrow \infty $, both the temperature and the ring radius approach the finite values $\frac{\sqrt{2}}{\pi }$ and $\frac{1}{2}$, respectively, indicating the stability of holographic material in this limit.

\paragraph{Non-periodic solution.}

Consider now the case of small dislocation parameters. We showed that the first three conditions that ensure the existence of physical solutions are satisfied in the positive branch of the torsion field when the dislocation parameter lies in the narrow interval $c_{02} \leq |c|<c_\infty$, see also Table~\eqref{tab:critical points}. In the negative branch, these conditions are satisfied when the dislocation parameter is in the interval $c_{02}\leq |c|<1$. In both cases, the sign of the torsion has to be chosen according to Eq.~\eqref{rho'}.

Using the obtained solutions, the torsion strength becomes
\begin{equation}
\Omega _{0,\pm}(c)=\sqrt{\frac{c^{2}}{3(1-c^{2}) }}\coth \left(m^{\frac{1}{4}} \,\omega_0(c)R(c) \right) \,.
\label{eq:nonperiodic-sol}
\end{equation}
It is always positive because the argument of the hyperbolic cotangent is always positive. Thus, according to \eqref{rho'}, it also has to be fulfilled that $1<\frac{3(1-c^{2})}{c^{2}}\, \Omega_{0,\pm} ^{2}$.

Since the l.h.s.~of the above equation does not depend on the mass parameter and the r.h.s.~does, 
to ensure consistency, the mass parameter has to depend on  the dislocation parameter, $c$, namely,
\begin{equation}
m=\frac{\mathrm{arccoth}^4\left( \sqrt{\frac{3\left(
1-c^{2}\right) }{c^{2}}}\,\Omega _{0,\pm}\right)}{\omega_0^4R^4}\,,
\label{mass-aperiodic}
\end{equation}
in a suitable range of the dislocation parameter $c$ for each branch of the torsion field.

\begin{figure}[t!]
    \centering
\includegraphics[width=\linewidth]{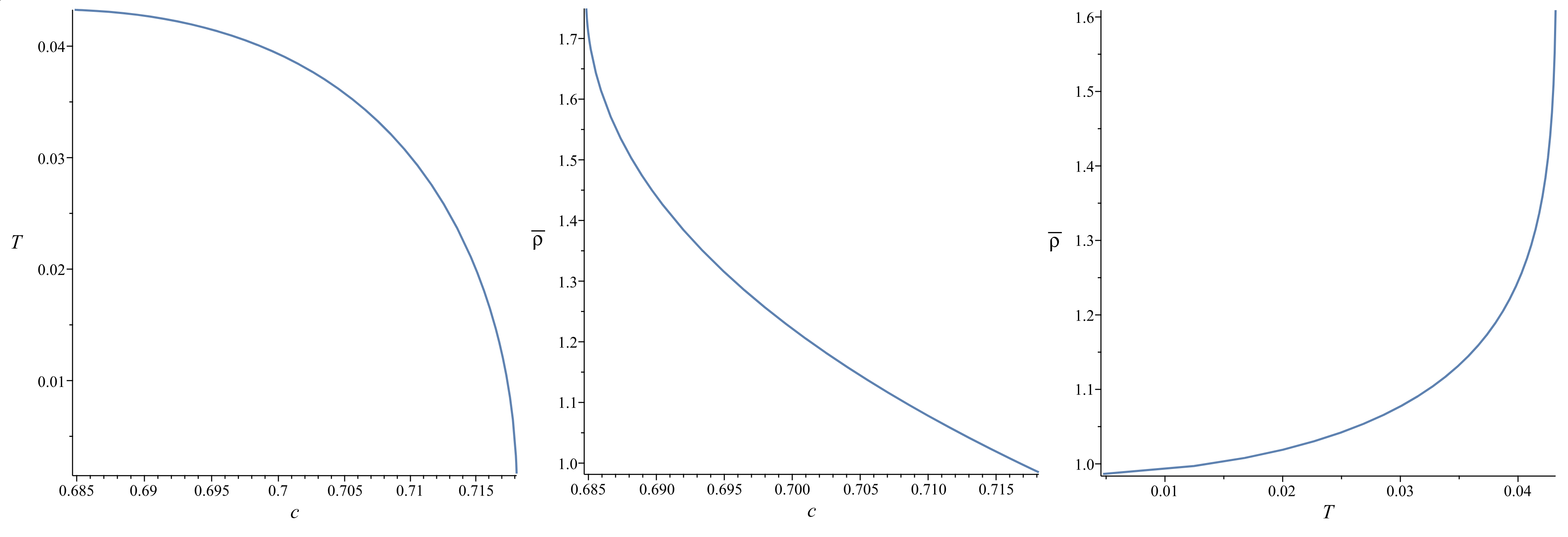}
    \caption{Temperature (left) and the ring radius (center) as functions of the dislocation parameter $c$, as well as the dependence $\bar\rho(T)$ (right), in case of the non-periodic solution, the positive branch. The solution exists in the narrow interval $c_{02} \leq |c| <c_\infty$, as shown in Table \eqref{tab:critical points}. }
    \label{FigP.nonperiodic}
\end{figure}

\begin{figure}[t!]
    \centering
\includegraphics[width=\linewidth]{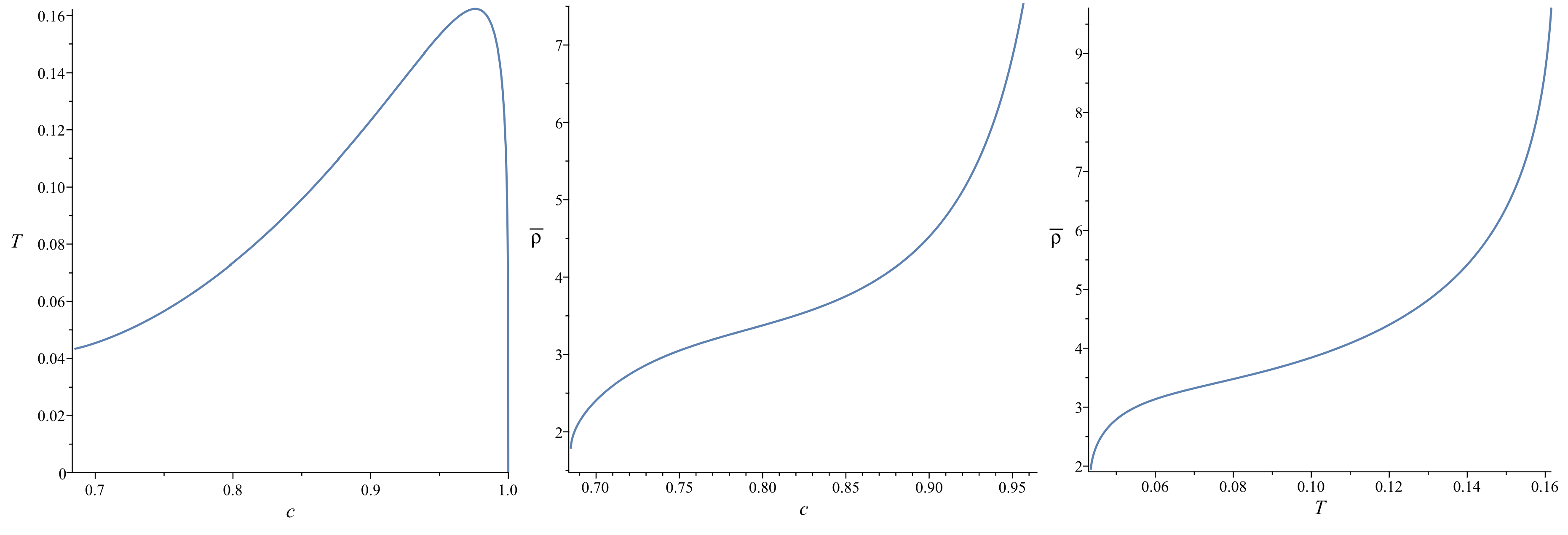}
    \caption{Temperature (left) and the ring radius (center) as functions of the dislocation parameter in the case of the non-periodic solution, the negative branch. The direct dependence $\bar\rho(T)$ (right). The solution exists in the interval $c_{02} \leq |c| <1$, as shown in Table \eqref{tab:critical points}.}
    \label{FigN.nonperiodic}
\end{figure}

Dependence of the temperature and the radius (see Eqs.~\eqref{bR} and \eqref{R}) from the dislocation parameter, for both positive and negative branches of the diagonal torsion, is shown in  Figs.~\ref{FigP.nonperiodic} and \ref{FigN.nonperiodic}, which also include numerically obtained form of the ring radius as a function of the temperature,  $\bar\rho(T)$.

For the positive branch, the crystalline material exists in a very narrow interval of the dislocation parameter, exhibiting significant changes in its properties. For instance, the temperature decreases monotonically, as illustrated in Fig.\ref{FigP.nonperiodic} (left), while Fig.\ref{FigP.nonperiodic}(center) shows that the radius reduces its size notably in this interval. Ultimately, the radius of the ring increases with rising temperature, as shown in Fig.~\ref{FigP.nonperiodic} (right).

In the negative branch, the temperature has a maximum, as shown in Fig.~\ref{FigN.nonperiodic}(left). It starts at the small but finite temperature close to $c_{02}$, grows until  $T_{\mathrm{max}}\approx 0.162$ at $|c|\approx 0.976$, and then rapidly falls off to small values, as $|c|$ approaches $1$.  The radius then monotonously increases, see  Fig.~\ref{FigN.nonperiodic}(center), starting at a small value for small $|c|$, and growing until large values as $|c|$ approaches $1$. On the other hand, as a function of the temperature, the radius increases from the very small values to the large ones, as shown in Fig.~\ref{FigN.nonperiodic}(right).  \bigskip

We can observe that the behavior of solutions is very different in the three described cases. For example, the radius of the ring gets smaller as the temperature rises in the periodic case, whereas it gets larger in the non-periodic case. This lack of universality results from the non-linearity of the underlying holographically dual theory, that is,  CS AdS gravity. Physically interesting cases are the ones where the system features dislocations with the torsion far from critical points, where either the temperature of the radius diverges, for instance, close to $\pm c_*$ for the periodic solution, or $\pm c_{02}$ and $\pm 1$ for the non-periodic ones.

The previous analysis applies to an ideal holographic QFT that does not possess dissipation that may emerge, for instance,  from additional matter fields in the theory. Including the dissipation might smooth some divergences occurring close to the critical torsion, i.e., dislocation configuration.

\subsection{Burgers vector}
\label{Burgers}

Previously, we obtained a solution for the holographic QFT at temperature $T$, featuring a dislocation localized on the ring $\mathcal{R}$, which
represents a defect of the material.  
More precisely, the axially symmetric torsion field is nontrivial along the circle $\mathcal{R}$ with the radius $\bar{\rho}$ located in the $\bar{z}=0$ plane. The observable quantities --the temperature and the ring radius-- depend on the dislocation parameter $c$, which describes the internal twisting of the torsion field, or the size of the torsional vortex, as well as the intensity $\Omega(c)$ of the torsion field. It exists only in three intervals of $c$ describing three different types of dislocations. Out of the ring line, the torsion field is zero, however, the temperature is still different than zero since the whole system is in thermal equilibrium. (This is possible because, on the gravity side, the black hole exists in the whole spacetime.) Therefore, $T$ is a continuous, constant parameter in the entire space. In contrast, the torsion field is not continuous because it possesses an $\mathbb{S}^1$ topological defect, the localized torsional vortex.

The contorsion field generated by the
dislocation takes the form
\begin{equation}
K^{ab}=\left\{
\begin{array}{lll}
\bar{K}^{ab}\,, & \rho =\bar{\rho}\,, & z=\bar{z}=0\,, \\
0\,, & \rho \neq \bar{\rho}\,, & z\neq 0\,.
\end{array}
\right.
\end{equation}
Using the metric \eqref{g,k}, \eqref{cyl} and the relation $A_{i}=8c\,B_{i}$
given by \eqref{f(c)}, we find in the points of the ring $\mathcal{R}$ that
the contorsion is
\begin{equation}
\bar{K}_{\mu \nu }=\left( -\frac{1}{32}\,\bar{\rho}\,\epsilon _{\mu \nu
\alpha \beta }\,A^{\beta }-B_{\mu }g_{\nu \alpha }+B_{\nu }g_{\mu \alpha
}\right) \mathrm{d}x^{\alpha }\,.
\end{equation}
It has six independent components. Taking into account that $A_{t},A_{\varphi }=0$ and $B_{t},B_{\varphi }=0$, and applying the notation 
\eqref{Levi 3D} from Appendix \ref{conventions}, namely $\epsilon _{t\rho
\varphi z}=1$, we find
\begin{equation}
\begin{array}{llll}
\bar{K}^{01} & =\Omega \left( \mathrm{d}t+4c\bar{\rho}\,\mathrm{d}\varphi
\right) \,,\medskip \qquad  & \bar{K}^{12} & =-\Omega \left( 4c\,\mathrm{d}t+%
\bar{\rho}\mathrm{d}\varphi \right) \,, \\
\bar{K}^{02} & =4c\Omega \,\left( -\mathrm{d}\rho +\mathrm{d}z\right)
\,,\medskip  & \bar{K}^{13} & =\Omega \left( \mathrm{d}\rho -\mathrm{d}
z\right) \,, \\
\bar{K}^{03} & =\Omega \left( \mathrm{d}t-4c\bar{\rho}\,\mathrm{d}\varphi
\right) \,, & \bar{K}^{23} & =\Omega \left( -4c\,\mathrm{d}t+\bar{\rho}
\mathrm{d}\varphi \right) \,.
\end{array}
\label{K}
\end{equation}
In the computation, we replaced the components of the metric. We used $%
B_{\rho }=B_{z}=\Omega $, as well as $A^{\rho }=A^{z}=32c\,\Omega $,
projecting also the four-dimensional spacetime indices to the tangent space
using the vielbein \eqref{e,k BH}, namely, $\bar{K}^{ab}=e^{a\mu }e^{b\nu }%
\bar{K}_{\mu \nu }$.

This defect is characterized by the Burgers vector. In the geometric theory of defects, the Burgers vector is the flux of the torsion tensor over a spatial surface, $b^{i}=\int T^{i}$ \cite{kleinert1989gauge}. Our system is Lorentz-covariant, such that we define a $4$-vector
\begin{equation}
b^{a}=\int\limits_{\mathcal{R}}T^{a}=\frac{1}{2}\,\int\limits_{\mathcal{R}%
}T_{\mu \nu }^{a}\,\mathrm{d}x^{\mu }\wedge \mathrm{d}x^{\nu }\,.
\end{equation}
Each component of the torsion tensor contains a (two-dimensional) Dirac delta
function restricting it to the value $\bar{T}_{\mu \nu }^{a}$, nontrivial on the ring $\mathcal{R}$, located at $\rho =\bar{\rho}$, $z=0$.
Because our configuration is static, we will evaluate the above integral at the space-like surface at constant time, taking
$\mathrm{d}t=0$. Then from $\bar{T}^{a}=\bar{K}^{ab}\wedge e_{b}$, the four-dimensional
contorsion \eqref{K} and the metric \eqref{e,k BH}, we find
\begin{eqnarray}
\bar{T}^{0} &=&4c\bar{\rho}\Omega \,\left( \mathrm{d}z-\mathrm{d}\rho
\right) \wedge \mathrm{d}\varphi\,,  \notag \\
\bar{T}^{1} &=&-\bar{T}^{3}=\frac{\Omega }{2}\,\mathrm{d}\rho \wedge \mathrm{d}z \,,  \label{barT}\\
\bar{T}^{2} &=&-\frac{\bar{\rho}\Omega }{2}\left( \mathrm{d}z+\mathrm{d}\rho
\right) \wedge \mathrm{d}\varphi \,,  \notag
\end{eqnarray}
which leads to the  Burgers 4-vector projected from the tangent space to the spacetime manifold $b^{\mu }=e_{a}^{\mu }b^{a}=(b^{t},\vec{b})$
with the components
\begin{equation}
b^{t}=-32\pi c\bar{\rho}\,\Omega \,,\qquad \vec{b}=\left(\Omega
,0,-\Omega  \right) \,.
\label{eq:Burgers-vector}
\end{equation}
The fact that it is non-vanishing means that there is a topological defect, or a torsional vortex, associated with the configuration of the torsion field. The torsion field strength $\Omega$ is also the defect's strength or the Burgers vector's size. It describes the axially-symmetric mixed screw-edge dislocation (because $b^{\rho }, b^{z} \neq 0$) in a purely spatial sector, i.e., without taking into account the time direction. However, the time scale also appears related to the $b^{t}$ component, suggesting the existence of the time crystal, where time-translational symmetry is spontaneously broken~\cite{Zaletel-2023}. 

To further elucidate the nature of the dislocation defects, we calculate the norm square of the Burgers vector \eqref{eq:Burgers-vector}, and find 
\begin{equation}
    b^{2}=\left( \frac{1}{2}-256\,\pi ^{2}c^{2}\bar{\rho}^{2}(c)\right) \Omega ^{2}(c)\,.
\end{equation}
It can be shown that $b^\mu$ is a space-like vector (with $b^2$ positive\footnote{We use the signature of the metric  $(-,+,+,+)$ so that $b^2>0$ corresponds to a  space-like vector.})
for the periodic solution in the very narrow range of the dislocation parameters $c_* < c < 1.171$ (where $c_* \approx 1.169$),  with the upper limit in this interval corresponding to the space-time crystal for which the Burgers vector is null, $b^2=0$. In all other allowed dislocation ranges, the Burgers vector has the time-like character, $b^2<0$, and therefore it corresponds to the time-like dislocation defect in time crystal, with the continuous time-translation symmetry spontaneously broken. The emerging time scale associated with the breaking of the time translation symmetry is then $\sim\sqrt{-b^2}$.   Properties of space-time supersolids have also been discussed using holography in Ref.~\cite{Yang:2023dvk}.
\medskip

Finally, we comment on the relationship between the torsional vortex \eqref{barT} and the bulk geometry, namely, CS AdS gravity. In the four-dimensional holographic QFT, this defect is a codimension-2 surface, i.e., it is a line in the 3D transversal plane.  On the gravity side, the components of torsion that contain defects affect both AdS curvature and bulk torsion, as can be seen from their radial expansions given in Eqs.~\eqref{Fexp}. Therefore, from the point of view of the gravitational theory, the topological defects appear as codimension-3 surfaces in the torsional field.

\section{Conclusions and discussion}
\label{Sec:Conclusions}

We obtained nontrivial solutions in five-dimensional Einstein-Gauss-Bonnet
gravity in AdS spacetime, with the Gauss-Bonnet coupling fixed at the
Chern-Simons point. The resulting  geometries correspond to black holes with finite
Hawking temperature, while the torsion field is localized along a ring. The configuration describes a ring-shaped topological defect characterized by the parameter $c$.

Our detailed analysis shows that there are two families of axially-symmetric solutions in such a theory for
which the torsion is either periodic, Eq.~\eqref{eq:periodic-sol}, with its
characteristic features shown in Fig.~\ref{Fig.periodic}, or aperiodic, as
given by Eq.~\eqref{eq:nonperiodic-sol}. The black hole mass $M=2mc^2$ in both cases is fully determined by the parameter $c$, see Eqs.~\eqref{eq:mass-periodic} and \eqref{mass-aperiodic}. The aperiodic class of solutions further exhibits positive and negative branches, displayed in Figs.~\ref{FigP.nonperiodic} and \ref{FigN.nonperiodic}, respectively. We point out
that the behavior of solutions is very different in these cases. For
example, the ring radius behaves oppositely for the two families of solutions: as the temperature increases, it decreases in the periodic case, whereas it increases in the non-periodic case. Such a nonuniversal behavior
emerges from the non-linearity of the underlying theory, that is, CS AdS
gravity.

In our approach, the torsion field (dislocation strength), the mass of the black hole (temperature), and the radius of the ring as a function of the dislocation parameter, $c$, are determined self-consistently. Physically interesting cases correspond to the dislocation parameter away from the critical values, where either the temperature or the ring radius diverges; for instance, such divergences occur close to $\pm c_*$ for the periodic solution, or $\pm c_{02}$ and $\pm 1$ for the non-periodic ones (see Table \eqref{tab:critical points}).

Using the AdS/CFT correspondence, the bulk solution can be understood from the perspective of the holographically dual theory. In particular, EGB AdS gravity is dual to a four-dimensional strongly coupled QFT at finite temperature, such as superconductors \cite{Jing:2010cx,Kanno:2011cs,Ge:2011cw} or insulators \cite{Pan:2011ah}, when coupled to scalar and electromagnetic fields.

In our case, the construction suggests that the holographic QFT on the boundary may represent a strongly coupled system at finite temperature with dislocations in the crystalline structure. Moreover, since such dislocations can produce an unequal number of chiral quasiparticles, there is a possibility that the Abelian anomaly $\mathcal{A}_{\mathrm{odd}}$ realized on the ring corresponds to a chiral anomaly, with the system in this case potentially describing holographic Weyl semimetals.

Indeed, both classes of solutions yield the odd-parity Abelian anomaly arising from the topological nontriviality of the QFT in the form of a ring-shaped defect. This anomaly is therefore a genuinely non-linear effect that cannot be treated perturbatively.

The anomaly can be computed using \eqref{dB} and \eqref{rho}, as
\begin{equation}
\mathcal{A}_{\mathrm{odd}}(c)=32^{2}\kappa mc^{3}\left[ \frac{m^{\frac{3}{4}}\Omega _{0,\pm }}{R}+\frac{1+2c^{2}}{2c^{2}}\,m\left( 1+\frac{3(c^{2}-1)}{c^{2}}\,\Omega _{0,\pm }^{2}\right) \right] ,
\label{eq:anomaly-on-solution}
\end{equation}
where $m$, $\Omega _{0,\pm }$ and $R$ are known functions of $c$. The anomaly exists only in the range of $c$ where these quantities are
well-defined. The nontriviality of $\mathcal{A}_{\mathrm{odd}}$ follows from the presence of the ring, such that it cannot be expressed globally as a total divergence. Moreover, it is proportional to
the Nieh-Yan invariant \cite{Nieh:1981ww},
\begin{equation}
T^{a}\wedge T_{a}-R^{ab}\wedge e_{a}\wedge e_{b}=\mathcal{J}_{\mathrm{NY}}\,%
\mathrm{d}^{4}x\,.
\end{equation}
Indeed, from the torsion decomposition \eqref{A,B} and the planar black hole
metric \eqref{g,k}, one finds
\begin{eqnarray}
T^{a}\wedge T_{a} &=&\frac{3\rho }{8}\,A^{\mu }B_{\mu }\,\mathrm{d}^{4}x\,,
\notag \\
R^{ab}\wedge e_{a}\wedge e_{b} &=&\frac{3}{16}\,\left( \rule[2pt]{0pt}{10pt}%
\partial _{\mu }\left( \rho A^{\mu }\right) +2\rho A^{\mu }B_{\mu }\right) \,
\mathrm{d}^{4}x\,.
\end{eqnarray}
Combining both terms yields
\begin{equation}
\mathcal{A}_{\mathrm{odd}}=\frac{8mc^{2}}{3\pi G\bar{\rho}}\,\mathcal{J}_{\mathrm{NY}}\,.
\label{eq:NY-anomaly-solution}
\end{equation}%
This Abelian anomaly receives a non-vanishing contribution only on the ring $\rho =\bar{\rho}$, with $m$ and $\bar{\rho}$ depending on $c$.
We find that, in four-dimensional QFT, the anomaly coefficient scales as $N^{2}\sim \frac{\ell ^{3}}{G}$, where $N$ is the effective number of degrees of freedom.
Torsion-induced chiral anomalies are also known to be proportional to the
Nieh-Yan invariant \cite{Chandia:1997hu,Chandia:1999az}, and renormalization
conditions may affect their coefficients \cite{Erdmenger:2024zty}. These results point that $\mathcal{A}_{\mathrm{odd}}$ could be a chiral anomaly, in which case the obtained holographic theory could contain the quasiparticles analogous to those in Weyl semimetals. Nevertheless, for a clear answer, further investigation is needed. A key open problem is the mechanism by which the $\mathrm{U}(1)$ symmetry in the bulk, inherited from the $\mathrm{SO}(4,2)$ gauge symmetry, is broken at the boundary, giving rise to the Abelian anomaly. The answer lies in the specific choice of boundary terms in the renormalized action. A similar mechanism leading to a 2D chiral anomaly has been illustrated in a toy model discussed in Ref.~\cite{Banados:2006fe}.

Holographic description of the dislocations motivates the study of its lattice version.
In particular, it would be interesting to find the lattice realization of the torsion field in terms of the dislocation configuration and its coupling to the fermions. This would permit to directly test the predictions of our holographic theory in a lattice model, particularly the dependence of the Burgers vector and anomaly on the dislocation parameter encoding the strength of the torsion.

A rather important pursuit in the holographic setup concerns the instabilities of the holograhic system, particularly by explicitly including extra gauge and matter fields in the bulk and investigating possible patterns of symmetry breaking encoding the interaction-driven instabilities on the lattice,  such as axion insulators \cite{Wang-PRB2013,Roy-Sau-2013,RGJ-PRB-2017,Gooth2019}. This problem can also be related to the fact that a supersymmetric extension in AdS space was found in five dimensions in \cite{Chamseddine:1990gk}, and also in higher-dimensional odd-dimensional spacetimes \cite{Banados:1996hi,Troncoso:1997va,  Troncoso:1998ng}, and it contains black hole solutions with nontrivial topological charges \cite{Andrianopoli:2021qli}. Using these theories, it may also be conceivable to construct a holographic dual of strongly coupled states on the surface of topological insulators that feature supersymmetric critical points in $3+1$ dimensions \cite{Lee-2007,Roy-PRB2013,Grover-2014,Ponte_2014}.

Other possible future lines of research include the use of different types of matter fields to introduce a dislocation in the QFT. Einstein-Gauss-Bonnet AdS gravity away from the CS point represents yet another prospect for future pursuits since it contains an additional free parameter for holographic modeling, the GB coupling constant.

\section*{Acknowledgments}

The authors would like to thank Diego Correa, Alberto Faraggi, Radouane Gannouji, Ayan Mukhopadhyay and Rodrigo Soto for their useful comments. This work has been funded in part by Anillo Grant ANID/ACT210100 \textit{Holography and its Applications to High Energy Physics, Quantum Gravity and Condensed Matter Systems} and FONDECYT Regular Grants 1230492, 1230933 (V.J.) and 1231779. This work is also supported by the Swedish Research Council Grant No.~VR 2019-04735 (V.J.), and the Chilean ANID BECAS/DOCTORADO NACIONAL scholarship No.~21211990 (F.R.) and the UTFSM scholarship (F.R.).   

\appendix

\section{Notation and conventions}
\label{conventions}
Here we define some conventions used in the main text.

\paragraph{Local coordinates.}

We consider the five-dimensional spacetime that locally has the form of the
cylinder $\mathcal{M}=\mathbb{R}\times \Sigma $, such that $\mathbb{R}$
corresponds to the time coordinate and $\Sigma $ is the spatial manifold at
constant time. Notation for the local coordinates, in both the manifold and
tangent spaces, are summarized in the following table:
\begin{equation*}
\frame{$%
\begin{array}{lll}
\text{\textbf{5D Lorentz indices:}}\hspace{.7cm} & \text{\textbf{4D Lorentz indices:}} &
\text{\textbf{3D Lorentz indices:}} \\
A=(a,4) & a=(0,m) & m=1,2,3 \\
A=0,1,2,3,4\medskip  & a=0,1,2,3 &  \\
\text{\textbf{5D spacetime: }} & \text{\textbf{4D boundary: }}\sigma =%
\mathrm{const.} & \text{\textbf{3D space: }}t,\sigma =\mathrm{const.} \\
\mathcal{M}=\mathbb{R}\times \Sigma  & \partial \mathcal{M}=\mathbb{R}\times
\partial \Sigma _{\infty } & \partial \Sigma  \\
x^{M}=(x^{\mu },\sigma )\medskip  & x^{\mu }=\left( t,x^{i}\right)  &
x^{i}\,,i=1,2,3 \\
\text{\textbf{5D black hole: }} & \text{\textbf{4D black-hole boundary:}}\hspace{.7cm} &
\text{\textbf{3D flat space: }} \\
x^{M}=\left( t,y^{i},\sigma \right)  & x^{\mu }=\left( t,y^{i}\right)  &
y^{i}=\left( \rho ,\varphi ,z\right)  \\
t\in \mathbb{R}\text{, }\sigma \geq 0 &  & \rho \geq 0\text{, }\varphi \in
[0,2\pi ]\text{, }z\in \mathbb{R}
\end{array}
$}
\end{equation*}
For three-dimensional indices, we use the Latin letters
\begin{equation}
\begin{array}[b]{ll}
\text{flat indices:\qquad } & m,n,s,p,q,\ldots\,,  \\
\text{curved indices:} & i,j,k,l,\ldots \,.
\end{array}
\end{equation}

\paragraph{Levi-Civita symbol.}

We also define the five- and four-dimensional volume elements as 
\begin{eqnarray}
\mathrm{d}x^M \wedge \mathrm{d}x^N\wedge \mathrm{d}x^K \wedge \mathrm{d}x^L \wedge \mathrm{d}x^S &=&-\mathrm{d}^5 x\,\epsilon ^{MNKLS}\,,  \notag \\
\mathrm{d}x^{\mu }\wedge \mathrm{d}x^{\nu }\wedge \mathrm{d}x^{\alpha}\wedge \mathrm{d}x^{\beta } &=&-\mathrm{d}^{4}x\,\epsilon ^{\mu \nu \alpha
\beta }\,,  \label{volume 4D} \\
\mathrm{d}t\wedge \mathrm{d}y^{i}\wedge \mathrm{d}y^{j}\wedge \mathrm{d}%
y^{k} &=&-\mathrm{d}^{4}x\,\epsilon ^{ijk}\,, \notag 
\end{eqnarray}
and the following notation for the constant Levi-Civita symbol:
\begin{eqnarray}
\epsilon _{abcd4} &=&\epsilon _{abcd},\qquad \epsilon _{tijk}\equiv \epsilon
_{ijk}  \notag \\
\epsilon _{tijk} &=&\epsilon _{ijk}\,,\qquad \epsilon ^{tijk}=-\epsilon
^{ijk}\,.  \label{Levi 3D}
\end{eqnarray}
Explicitly, in cylindrical coordinates, we use the conventions $\epsilon _{t\rho \varphi z}=1$ and $\epsilon _{\rho \varphi z}=1$.

\paragraph{AdS algebra.}

In tangent space, the flat metric is mostly positive, $\eta _{AB}=%
\mathrm{diag}(-,+,+,+,+)$. The usual Lorentz isometries are extended to
anti-de Sitter (AdS) isometries in Chern-Simons AdS gravity.\
Five-dimensional AdS algebra, isomorphic to $\mathfrak{so}(2,4)$, can be
written in the basis of Lorentz rotations $J_{AB}=-J_{BA}$ and the AdS
translations $P_{A}$ as
\begin{eqnarray}
\left[ J_{AB},J_{CD}\right]  &=&\eta _{AD}J_{BC}-\eta _{BD}J_{AC}-\eta
_{AC}J_{BD}+\eta _{BC}J_{AD},  \notag \\
\left[ J_{AB},P_{C}\right]  &=&-\eta _{AC}P_{B}+\eta _{BC}P_{A}\,,\qquad %
\left[ P_{A},P_{B}\right] =J_{AB}\,.  \label{AdS algebra 1}
\end{eqnarray}%
Using the decomposition of Lorentz indices $A=(a,4)$ and the fact that $\eta
_{44}=1$ and $\eta _{ab}=\mathrm{diag}(-,+,+,+)$, the above algebra can be
rewritten in the basis $\left\{ J_{ab},J_{a}^{\pm }\right\} $, where $%
J_{a}^{\pm }\equiv P_{a}\pm J_{a4}$, as%
\begin{eqnarray}
\left[ J_{ab},J_{cd}\right]  &=&\eta _{ad}J_{bc}-\eta _{bd}J_{ac}-\eta
_{ac}J_{bd}+\eta _{bc}J_{ad}\,,  \notag \\
\left[ J_{ab},J_{c}^{\pm }\right]  &=&-\eta _{ac}J_{b}^{\pm }+\eta
_{bc}J_{a}^{\pm }\,,  \label{AdS algebra 2} \\
\left[ J_{a}^{+},J_{b}^{-}\right]  &=&2J_{ab}-2\eta _{ab}P_{4}\,,\qquad %
\left[ J_{a}^{\pm },P_{4}\right] =\pm J_{a}^{\pm }\,,  \notag
\end{eqnarray}%
where all other commutators are zero.

\paragraph{Isometries.}

Five-dimensional static, spherical planar black hole has isometries given by the translations
$p_{m}$  and rotations $j_{m}$ in the 3D horizon plane, as well as temporal
translations $p_{0}$, given in cylindrical coordinates by%
\begin{equation}
\begin{array}{llll}
p_{0} & =\partial _{t}\,, &  &  \\
p_{1} & =\cos \varphi \,\partial _{\rho }-\dfrac{\sin \varphi }{\rho }%
\,\partial _{\varphi }\,,\quad \medskip  & j_{1} & =-z\sin \varphi
\,\partial _{\rho }-\dfrac{z}{\rho }\,\cos \varphi \,\partial _{\varphi
}+\rho \,\sin \varphi \,\partial _{z}\,, \\
p_{2} & =\sin \varphi \,\partial _{\rho }+\dfrac{\cos \varphi }{\rho }%
\,\partial _{\varphi }\,,\medskip  & j_{2} & =-z\cos \varphi \,\partial
_{\rho }+\dfrac{z}{\rho }\,\sin \varphi \,\partial _{\varphi }+\rho \,\cos
\varphi \,\partial _{z}\,, \\
p_{3} & =\partial _{z}\,, & j_{3} & =\partial _{\varphi }\,.%
\end{array}
\label{Killing}
\end{equation}%
They satisfy the Lie-bracket algebra $\mathrm{ISO}(3)$%
\begin{equation}
\lbrack j_{m},j_{n}]=-\epsilon _{mnk}\,j_{k}\,,\qquad \lbrack
j_{m},p_{n}]=\epsilon _{mnl}\,p_{k}\,,\qquad \lbrack p_{m},p_{n}]=0\,.
\label{Kalg}
\end{equation}

\paragraph{Axial and diagonal torsion. }

Axial torsion field $A^{\mu }(x)$ and diagonal torsion field $B_{\mu }(x)$
are 4-vectors. However, in the axially-symmetric ansatz used from Subsec.~\ref{Dislocation}, where the temporal components vanish identically and the
fields are static, $A^{i}(y)$ and $B_{i}(y)$ become 3-vectors defined in the
transversal section $\partial \Sigma $. There the metric $\gamma _{ij}$ and
its inverse $\gamma ^{ij}$ lower and rise the spatial indices. We also use
tilde to emphasize that the quantity is treated as three-dimensional, using
the following notation:
\begin{equation*}
\fbox{$
\begin{array}{lll}
\mathrm{d}\sigma _{i}=\dfrac{\rho }{2}\,\epsilon _{ijk}\,\mathrm{d}%
y^{j}\wedge \mathrm{d}y^{k}\,,\medskip \quad  & \mathrm{d}\tilde{\sigma}_{m}=%
\tilde{e}_{m}^{i}\,\mathrm{d}\sigma _{i}\,, &  \\
A_{ij}=\rho \,\epsilon _{ijk}\,A^{k},\medskip  & \tilde{A}_{\ j}^{m}=\tilde{e}^{mi}A_{ij}\,, & A_i=\gamma_{ij}A^j\,, \\
\tilde{A}^{mn}=\tilde{e}^{mi}\tilde{e}^{nj}A_{ij},\medskip  & \tilde{A}=%
\dfrac{1}{2}\,A_{ij}\,\mathrm{d}y^{i}\wedge \mathrm{d}y^{j}\,, &  \\
\tilde{A}^{m}=\tilde{A}_{\ i}^{m}\mathrm{d}y^{i},\medskip  & \mathcal{A}^{m}=%
\tilde{e}_{i}^{m}A^{i}\,, & \mathcal{A}^{2}=\gamma _{ij}A^{i}A^{j}\,, \\
\tilde{B}^{m}=\tilde{e}^{mi}B_{i}\,,\medskip  & \tilde{B}=B_{i}\mathrm{d}
y^{i}\,, & B^{2}=\gamma ^{ij}B_{i}B_{j}\,,
\end{array}%
$}
\end{equation*}
which results in the identities
\begin{equation}
\tilde{A}^{m}\wedge \tilde{e}_{m}=-2\tilde{A}\,,\qquad \tilde{A}^{mn}\,%
\tilde{e}_{n}=\tilde{A}^{m}\,,\qquad \tilde{B}^{n}\tilde{e}_{n}=\tilde{B}\,.
\end{equation}

\section{Torsionless spin connection in five dimensions}
\label{Torsionless}


The five-dimensional torsionless (Levi-Civita) spin connection $\hat{%
\mathring{\omega}}^{AB}$ depends only on the vielbein $\hat{e}^{A}$, and it
is computed from $\mathrm{\hat{D}}\hat{e}^{A}=\mathrm{d}\hat{e}^{A}+\hat{%
\mathring{\omega}}^{AB}\wedge \hat{e}_{B}=0$. The solution in terms of the
Christoffel symbols,
\begin{equation}
\hat{\Gamma}_{KL}^{M}=\frac{1}{2}\,\hat{g}^{MN}\left( \partial _{K}\hat{g}%
_{NL}+\partial _{L}\hat{g}_{NK}-\partial _{N}\hat{g}_{KL}\right) \,,
\end{equation}%
is given by%
\begin{equation}
\hat{\mathring{\omega}}^{AB}=\hat{e}^{BM}\left( -\partial _{N}\hat{e}%
_{M}^{A}+\hat{\Gamma}_{NM}^{K}\,\hat{e}_{K}^{A}\right) \,\mathrm{d}x^{N}\,.
\end{equation}%
For the dimensionally continued black hole metric \eqref{BHmetric}, with
the planar horizon, the non-zero components of the Christoffel
symbols are
\begin{equation}
\begin{array}{llllll}
\hat{\Gamma}_{\sigma \sigma }^{\sigma } & =-\dfrac{1}{\sigma }\,, & \hat{%
\Gamma}_{zz}^{\sigma }= & -\dfrac{M^{2}\sigma ^{2}-1}{2}\,, & \hat{\Gamma}%
_{\sigma z}^{z} & =\dfrac{M\sigma -1}{2\sigma \left( M\sigma +1\right) }%
\,,\medskip  \\
\hat{\Gamma}_{\rho \rho }^{\sigma } & =-\dfrac{M^{2}\sigma ^{2}-1}{2}\,, &
\hat{\Gamma}_{\sigma t}^{t} & =\dfrac{M\sigma +1}{2\sigma \left( M\sigma
-1\right) }\,,\qquad  & \hat{\Gamma}_{\varphi \varphi }^{\rho } & =-\rho
\,,\medskip  \\
\hat{\Gamma}_{\varphi \varphi }^{\sigma } & =-\rho ^{2}\,\dfrac{M^{2}\sigma
^{2}-1}{2}\,,\qquad  & \hat{\Gamma}_{\sigma \rho }^{\rho } & =\dfrac{M\sigma
-1}{2\sigma \left( M\sigma +1\right) }\,, & \hat{\Gamma}_{\rho \varphi
}^{\varphi } & =\dfrac{1}{\rho }\,.\medskip  \\
\hat{\Gamma}_{tt}^{\sigma } & =\dfrac{M^{2}\sigma ^{2}-1}{2\ell ^{2}}\,, &
\hat{\Gamma}_{\sigma \varphi }^{\varphi } & =\dfrac{M\sigma -1}{2\sigma
\left( M\sigma +1\right) }\,, &  &
\end{array}%
\end{equation}%
As a result, the five-dimensional Levi-Civita connection has non-zero
components%
\begin{equation}
\hat{\mathring{\omega}}^{04}=\dfrac{1+M\sigma }{2\ell \sqrt{\sigma }}\,%
\mathrm{d}t\,,\qquad \hat{\mathring{\omega}}^{m4}=\dfrac{1-M\sigma }{2\sqrt{%
\sigma }}\,\tilde{e}^{m}\,,\qquad \hat{\mathring{\omega}}^{12}=-\mathrm{d}%
\varphi \,.
\end{equation}

Comparing with the general radial FG expansion \eqref{expansion} and the
four-dimensional fields in the black hole solution \eqref{e,k BH}, the
components $\hat{\mathring{\omega}}^{a4}$ are in agreement.

The associated Riemann tensor, $\hat{\mathring{R}}^{AB}=\mathrm{d}\hat{%
\mathring{\omega}}^{AB}+\hat{\mathring{\omega}}^{AC}\wedge \hat{\mathring{%
\omega}}_{C}^{\ \ B}$, has components
\begin{equation}
\begin{array}{llll}
\hat{\mathring{R}}^{0m} & =\dfrac{M^{2}\sigma ^{2}-1}{4\ell \sigma }\,%
\mathrm{d}t\wedge \tilde{e}^{m}\,,\qquad \qquad & \hat{\mathring{R}}^{04} & =%
\dfrac{M\sigma -1}{4\ell \sqrt{\sigma }\sigma }\,\mathrm{d}\sigma \wedge
\mathrm{d}t\,, \\
\hat{\mathring{R}}^{m4} & =-\dfrac{M\sigma +1}{4\sqrt{\sigma }\sigma }\,
\mathrm{d}\sigma \wedge \tilde{e}^{m}\,, & \hat{\mathring{R}}^{mn} & =-
\dfrac{\left( M\sigma -1\right) ^{2}}{4\sigma }\,\tilde{e}^{m}\wedge \tilde{e%
}^{n}\,.%
\end{array}%
\end{equation}

\section{`No-go' solutions for the dislocations}
\label{Computations}

In this section, we explicitly write some technical details in solving Eqs.~\eqref{C}, necessary for understanding the results presented in Sec.~\ref{WSM}.

\subsection{General identities}

The building blocks of the holographic equations are the following
differential forms, expressed in terms of the quantities defined in the last
part of Sec.~\ref{conventions}:

\begin{itemize}
\item 1-form $K^{ab}=e^{a\mu}e^{b\nu}K_{\mu\nu\lambda}\de x^\lambda$
\begin{eqnarray}
K^{0m} &=&\frac{1}{\ell}\,\tilde{B}^{m}\,\mathrm{d}t+\frac{\ell^2}{8}\,%
\tilde{A}^{m}+\ell B_{t}\,\tilde{e}^{m}\,,  \label{Kab} \\
K^{mn} &=& \frac{\ell}{8}\,\left(-\tilde{A}^{mn}\mathrm{d}t+A^{t}\,\epsilon
^{mnk}\tilde{e}_{k}\right)-\tilde{B}^{m}\tilde{e}^{n}+\tilde{B}^{n}\tilde{e}%
^{m}\,;  \notag
\end{eqnarray}

\item 2-form $\mathrm{D}k^{a}$
\begin{eqnarray}
\mathrm{D}k^{0} &=& \frac{M}{2}\,\de t\wedge \tilde{B}-\frac{\ell^3 M}{8}\,%
\tilde{A} \,,  \notag \\
\mathrm{D}k^{m} &=&\frac{\mathbf{\ell }M}{2}\,\left( B_{t}\,\mathrm{d}%
t\wedge \tilde{e}^{m}-\frac{\ell}{4}\,A^{t}\,\mathrm{d}\tilde{\sigma}^{m}+%
\tilde{e}^{m}\wedge \tilde{B}\right) ;  \label{Dka}
\end{eqnarray}

\item 2-form $T^{a}=\frac{1}{2}\,e^{a\mu}T_{\mu\alpha\beta}\,\de x^\alpha
\wedge \de x^\beta$
\begin{eqnarray}
T^{0} &=&\frac{1}{2}\,\mathrm{d}t\wedge \tilde{B}-\frac{\ell^3}{8}\,\tilde{A}%
\,,  \notag \\
T^{m} &=&-\mathrm{d}t\wedge \left( \frac{\ell^2}{8}\,\tilde{A}^{m}+\frac{\ell%
}{2}\,B_{t}\,\tilde{e}^{m}\right) -\frac{\ell^2}{8}\,A^{t}\,\mathrm{d}\tilde{%
\sigma}^{m}-\frac{\mathbf{\ell }}{2}\, \tilde{B}\wedge \tilde{e}^{m};
\label{Ta}
\end{eqnarray}

\item 2-form $F^{ab}$ with $A_{t}=0$ y $B^{t}=0$
\begin{eqnarray}
F^{0m} &=&\frac{1}{\ell }\,\mathrm{d}t\wedge \left( -\mathrm{\mathring{D}}%
\tilde{B}^{m}+\tilde{B}^{m}\tilde{B}-B^{2}\tilde{e}^{m}-\frac{\ell ^{4}}{64}%
\,\tilde{A}^{mn}\tilde{A}_{n}\right)   \notag \\
&&+\frac{\ell ^{2}}{8}\,\left( \mathrm{\mathring{D}}\tilde{A}^{m}-2\tilde{A}%
\,\tilde{B}^{m}-\tilde{B}^{n}\tilde{A}_{n}\wedge \tilde{e}^{m}\right) \,,
\notag \\
F^{mn} &=&\frac{\ell }{8}\,\mathrm{d}t\wedge \left[ \mathrm{\mathring{D}}%
\tilde{A}^{mn}+2\tilde{B}^{m}\tilde{A}^{n}-2\tilde{B}^{n}\tilde{A}%
^{m}+\left( \tilde{A}^{ms}\tilde{e}^{n}-\tilde{A}^{ns}\tilde{e}^{m}\right)
\tilde{B}_{s}\right]   \label{F} \\
&&-\mathrm{\mathring{D}}\tilde{B}^{m}\wedge \tilde{e}^{n}+\mathrm{\mathring{D%
}}\tilde{B}^{n}\wedge \tilde{e}^{m}+\left( -\tilde{B}^{m}\tilde{e}^{n}+%
\tilde{B}^{n}\tilde{e}^{m}\right) \wedge \tilde{B}  \notag \\
&&+\frac{\ell ^{4}}{64}\,\tilde{A}^{m}\wedge \tilde{A}^{n}+\left(
M-B^{2}\right) \,\tilde{e}^{m}\wedge \tilde{e}^{n}\,;  \notag
\end{eqnarray}
\end{itemize}

\subsection{Torsion field without diagonal component, \texorpdfstring{$B_{\mu}=0$}{B=0}}
\label{Only A}

We will show that the holographic dislocation does not exist without the field $B_\mu$. To prove it, let us assume
that $B_{\mu }=0$, such that the only non-vanishing component is the axial
torsion is $A_{\mu }\neq 0$.

In this case, the building blocks of the holographic equations \eqref{C}
have simpler form
\begin{equation}
\begin{array}{llll}
K^{0m} & =\dfrac{\ell ^{2}}{8}\,\tilde{A}^{m}\,,\qquad \qquad  & K^{mn} & =%
\dfrac{\ell }{8}\left( -\tilde{A}^{mn}\mathrm{d}t+A^{t}\,\epsilon ^{mnk}%
\tilde{e}_{k}\right) \,,\medskip  \\
\mathrm{D}k^{0} & =-\dfrac{\ell ^{3}M}{8}\,\tilde{A}\,, & \mathrm{D}k^{m} &
=-\dfrac{\ell ^{2}M}{8}\,A^{t}\,\mathrm{d}\tilde{\sigma}^{m},\medskip  \\
T^{0} & =-\dfrac{\ell ^{3}}{8}\,\tilde{A}\,, & T^{m} & =-\dfrac{\ell ^{2}}{8}%
\left( \mathrm{d}t\wedge \tilde{A}^{m}+A^{t}\,\mathrm{d}\tilde{\sigma}%
^{m}\right) \,,%
\end{array}%
\end{equation}%
and the 2-form $F^{ab}$ has the components,
\begin{eqnarray}
F^{0m} &=&\dfrac{\ell ^{2}}{8}\,\mathrm{\mathring{D}}\tilde{A}^{m}+\dfrac{%
\ell ^{3}}{64}\,\left( \tilde{A}^{mn}\tilde{A}_{n}\wedge \mathrm{d}%
t-A^{t}\,\epsilon ^{mns}\tilde{A}_{n}\wedge \tilde{e}_{s}\right) \,,  \notag
\\
F^{mn} &=&\dfrac{\ell }{8}\left( -\mathrm{\mathring{D}}\tilde{A}^{mn}\wedge
\mathrm{d}t+\mathrm{d}A^{t}\wedge \epsilon ^{mns}\tilde{e}_{s}\right) +%
\dfrac{\ell ^{4}}{64}\,\tilde{A}^{m}\wedge \tilde{A}^{n}  \notag \\
&&+\dfrac{\ell ^{2}}{64}\left( -\tilde{A}_{\ p}^{n}\epsilon ^{mps}+\tilde{A}%
_{\ p}^{m}\epsilon ^{nps}\right) A^{t}\,\mathrm{d}t\wedge \tilde{e}%
_{s}+\left( M-\dfrac{\ell ^{2}}{64}\,(A^{t})^{2}\right) \,\tilde{e}%
^{m}\wedge \tilde{e}^{n}\,.
\end{eqnarray}%
Notice that $\mathrm{\mathring{D}}\tilde{A}^{m}=\mathrm{d}\tilde{A}^{m}-%
\mathrm{d}\varphi \,\delta _{12}^{mn}\tilde{A}_{n}$, and similarly for $%
\mathrm{\mathring{D}}\tilde{A}^{mn}$.

Now we will solve the equations \eqref{C}. We first analyze the equation $%
C_{ab}=0$. The transversal components
\begin{equation}
C_{mn}=-\dfrac{\ell ^{5}M}{32}\,\epsilon _{mns}\,\mathrm{d}t\wedge \tilde{A}%
^{s}\wedge \tilde{A}=-\mathrm{d}^{4}x\,\frac{1}{2}\,\rho ^{2}\tilde{e}%
^{si}\epsilon _{mns}\epsilon _{ijl}A^{j}A^{l}=0\,,
\end{equation}%
vanish identically due to symmetry reasons ($\epsilon _{ijl}A^{j}A^{l}\equiv
0$), and we are left to solve only
\begin{equation}
C_{0m}=\dfrac{\ell ^{4}M}{32}\,A^{t}\epsilon _{mns}\,\mathrm{d}t\wedge
\tilde{A}^{n}\wedge \mathrm{d}\tilde{\sigma}^{s}\propto A^{t}\mathcal{A}%
_{m}=0\quad \Rightarrow \quad A^{t}=0\,,
\end{equation}%
if we want $A^{i}\neq 0$.

The next equation to analyze is $\bar{C}_{a}=0$. While the component $\bar{C}%
_{0}=0$ vanishes due to $\mathrm{D}k^{m}=0$, for other components we get
\begin{equation}
\bar{C}_{m}=\dfrac{\ell ^{4}M}{64}\,\epsilon _{mns}\,\mathrm{d}t\wedge
\tilde{A}\wedge \mathrm{\mathring{D}}\tilde{A}^{ns}=-\mathrm{d}^{4}x\,\frac{%
\ell ^{4}M}{34}\rho \,A^{i}\mathrm{\mathring{D}}_{i}\mathcal{A}_{m}\,,
\notag
\end{equation}%
which gives three differential equations
\begin{eqnarray}
\bar{C}_{1} &=&0\quad \Rightarrow \quad A^{i}\partial _{i}A^{\rho }-\rho
(A^{\varphi })^{2}=0\,,  \notag \\
\bar{C}_{2} &=&0\quad \Rightarrow \quad A^{i}\partial _{i}\left( \rho
A^{\varphi }\right) +A^{\varphi }A^{\rho }=0\,, \\
\bar{C}_{3} &=&0\quad \Rightarrow \quad A^{i}\partial _{i}A^{z}=0\,,  \notag
\end{eqnarray}%
where $A^{i}\partial _{i}=A^{\rho }\partial _{\rho }+A^{z}\partial _{z}$.
From the last equation, we distinguish the possibilities of the horizontal axial torsion, $A_{z}=0$, and the vertical axial torsion, $A_{z}\neq 0$.

\subsubsection{Horizontal axial torsion \texorpdfstring{$A_{z}=0$}{Az=0}}
\label{Horizontal}

If the axial torsion $A_{i}$ has only the horizontal components, $A_{z}=0$,
the equations are reduced to
\begin{eqnarray}
0 &=&\frac{1}{2}\,\partial _{\rho }\left( A^{\rho }\right) ^{2}-\rho
(A^{\varphi })^{2}\,,  \notag \\
0 &=&A^{\rho }\left[ \partial _{\rho }\left( \rho A^{\varphi }\right)
+A^{\varphi }\right] \,.
\end{eqnarray}%
If $A^{\rho }=0$, the first equation implies $A^{\varphi }=0$, leading to a trivial solution, without a dislocation. Thus, we need $A^{\rho }\neq 0$, when the equations can be solved as
\begin{equation}
A^{\rho }=\sqrt{C^{2}-\frac{Z^{2}(z)}{\rho ^{2}}}\,,\qquad A^{\varphi }(\rho
)=\frac{Z(z)}{\rho ^{2}}\,,\qquad A_{z}=0\,,  \label{pre-sol}
\end{equation}%
where $C\neq 0$ is an integration constant and $\zeta (z)$ is an arbitrary, real function that remains to be determined. Note that $A^{\rho }$ is
well-defined only if $|Z(z)|\leq |C|\rho $, so we need to carefully analyze
the geometry of the solution; for example, if $\rho =0$ is allowed, we need $%
Z=0$ and the solution is $A^{\rho }=C$. The other possibility is to impose
the existence of the minimum value $\rho _{\min }\neq 0$, which allows $%
Z\neq 0$.

Furthermore, the $C_{a}=0$ becomes
\begin{eqnarray}
C_{0} &=&-\dfrac{\ell ^{2}}{8}\,\epsilon _{mns}\,\mathrm{d}t\wedge \left(
\dfrac{\ell ^{4}}{64}\,\tilde{A}^{m}\wedge \tilde{A}^{n}+M\,\tilde{e}%
^{m}\wedge \tilde{e}^{n}\right) \wedge \tilde{A}^{s}=0\,,  \notag \\
C_{m} &=&\dfrac{\ell ^{4}}{64}\,\epsilon _{mns}\,\mathrm{d}t\wedge \left( 2\,%
\mathrm{\mathring{D}}\tilde{A}^{n}\wedge \tilde{A}^{s}+\mathrm{\mathring{D}}%
\tilde{A}^{ns}\wedge \tilde{A}\right) =0\,.
\end{eqnarray}%
It can also be written as
\begin{eqnarray}
C_{0} &\propto &\epsilon _{mns}\,\left( \dfrac{\ell ^{4}}{64}\,\tilde{A}_{\
i}^{m}\tilde{A}_{\ j}^{n}+M\,\tilde{e}_{i}^{m}\tilde{e}_{j}^{n}\right)
\tilde{A}_{\ l}^{s}\epsilon ^{ijl}\,,  \notag \\
C_{m} &\propto &\epsilon _{mns}\,\left( 2\,\mathrm{\mathring{D}}_{i}\tilde{A}%
_{\ j}^{n}\tilde{A}_{\ l}^{s}+\frac{1}{2}\,\mathrm{\mathring{D}}_{i}\tilde{A}%
^{ns}A_{jl}\right) \epsilon ^{ijl}\,.
\end{eqnarray}%
The component $C_{0}$ is identically zero due to symmetry. The second
equation gives
\begin{equation}
C_{m}\propto \mathrm{\mathring{D}}_{n}\mathcal{A}^{n}\mathcal{A}_{m}+2A^{i}%
\mathrm{\mathring{D}}_{i}\mathcal{A}_{m}=0\,,
\end{equation}%
applying the shorthand notation $\tilde{e}_{m}^{i}\mathrm{\mathring{D}}_{i}=%
\mathrm{\mathring{D}}_{m}$. Then, using the definition of the covariant
derivative, $\mathrm{\mathring{D}}_{i}\mathcal{A}_{m}=\partial _{i}\mathcal{A%
}_{m}+\mathring{\omega}_{im\tilde{n}}\,\mathcal{A}^{\tilde{n}}$, implies
that the components have the form
\begin{eqnarray}
\mathrm{\mathring{D}}_{i}\mathcal{A}_{1} &=&\partial _{i}A^{\rho }-\delta
_{i}^{\varphi }\rho A^{\varphi }\,,  \notag \\
\mathrm{\mathring{D}}_{i}\mathcal{A}_{2} &=&\partial _{i}\left( \rho
A^{\varphi }\right) +\delta _{i}^{\varphi }A^{\rho }\,,  \notag \\
\mathrm{\mathring{D}}_{i}\mathcal{A}_{3} &=&\partial _{i}A^{z}\,,
\end{eqnarray}%
and we also find
\begin{equation}
\tilde{e}_{n}^{i}\mathrm{\mathring{D}}_{i}\mathcal{A}^{n}=\partial _{\rho
}A^{\rho }+\partial _{z}A^{z}+\frac{1}{\rho }\,A^{\rho }\,.
\end{equation}%
Thus, the equations become
\begin{eqnarray}
C_{1} &=&0:\quad \left( \partial _{\rho }A^{\rho }+\partial _{z}A^{z}+\frac{1%
}{\rho }\,A^{\rho }\right) A^{\rho }+2A^{i}\partial _{i}A^{\rho }-2\rho
(A^{\varphi })^{2}=0\,,  \notag \\
C_{2} &=&0:\quad \left( \partial _{\rho }A^{\rho }+\partial _{z}A^{z}+\frac{1%
}{\rho }\,A^{\rho }\right) \rho A^{\varphi }+2A^{i}\partial _{i}\left( \rho
A^{\varphi }\right) +2A^{\varphi }A^{\rho }=0\,,  \notag \\
C_{3} &=&0:\quad \left( \partial _{\rho }A^{\rho }+\partial _{z}A^{z}+\frac{1%
}{\rho }\,A^{\rho }\right) A^{z}+2A^{i}\partial _{i}A^{z}=0\,.
\end{eqnarray}%
Plugging in the obtained solution \eqref{pre-sol}, and knowing that $A^{z}=0$
and $\partial _{\varphi }=0$, the equation $C_{3}$ cancels out and the other
two equations yield
\begin{eqnarray}
0 &=&\frac{3}{2}\,\partial _{\rho }(A^{\rho })^{2}+\frac{1}{\rho }\,(A^{\rho
})^{2}-2\rho (A^{\varphi })^{2}=\frac{C^{2}}{\rho }\,,  \notag \\
0 &=&\rho A^{\varphi }\partial _{\rho }A^{\rho }+5A^{\rho }A^{\varphi
}+2\rho A^{\rho }\partial _{\rho }A^{\varphi }=\frac{ZC^{2}}{\rho ^{2}\sqrt{%
C^{2}-\frac{Z^{2}}{\rho ^{2}}}}\,.
\end{eqnarray}%
It can be seen that the equations are satisfied only if the integration
constant is $C=0$, finally giving
\begin{equation}
A^{\rho }=\sqrt{-\frac{Z^{2}(z)}{\rho ^{2}}}\,,\qquad A^{\varphi }=\frac{Z(z)%
}{\rho ^{2}}\,,\qquad A_{z}=0\,,  \label{sol}
\end{equation}%
and a complex field $A^{\mu }$. For the last equation, we find
\begin{equation}
C=\frac{\ell ^{3}}{16}\,\epsilon _{mns}\mathrm{d}t\wedge \left[ \mathrm{%
\mathring{D}}\tilde{A}^{m}\wedge \mathrm{\mathring{D}}\tilde{A}^{ns}-\tilde{A%
}^{mq}\tilde{A}_{q}\wedge \left( \dfrac{\ell ^{4}}{64}\,\tilde{A}^{n}\wedge
\tilde{A}^{s}+M\,\tilde{e}^{n}\wedge \tilde{e}^{s}\right) \right] .
\end{equation}%
We can analyze term by term and apply the identity $[\mathrm{\mathring{D}}%
_{i},\mathrm{\mathring{D}}_{j}]V^{m}=\mathring{R}_{ij}^{mn}V_{n}=0$. We
obtain
\begin{eqnarray}
\epsilon _{mns}\mathrm{d}t\wedge \mathrm{\mathring{D}}\tilde{A}^{m}\wedge
\mathrm{\mathring{D}}\tilde{A}^{ns} &=&-2\mathrm{d}^{4}x\,\partial _{i}\left[
\left( \rho \tilde{e}^{mi}A^{j}-\rho \tilde{e}^{mj}A^{i}\right) \mathrm{%
\mathring{D}}_{j}\mathcal{A}_{m}\right] ,  \notag \\
-\epsilon _{mns}\mathrm{d}t\wedge \tilde{A}^{mq}\tilde{A}_{q}\wedge \dfrac{%
\ell ^{4}}{64}\,\tilde{A}^{n}\wedge \tilde{A}^{s} &=&0\,, \\
-\epsilon _{mns}\mathrm{d}t\wedge \tilde{A}^{mq}\tilde{A}_{q}M\,\tilde{e}%
^{n}\wedge \tilde{e}^{s} &=&-4\mathrm{d}^{4}x\,M\rho \,\mathcal{A}^{2}\,.
\notag
\end{eqnarray}%
Therefore,
\begin{equation}
C\propto \partial _{i}\left[ \rho \left( \tilde{e}^{mi}A^{j}-\tilde{e}%
^{mj}A^{i}\right) \mathrm{\mathring{D}}_{j}\mathcal{A}_{m}\right] +2M\rho \,%
\mathcal{A}^{2}=0\,,
\end{equation}%
which is equivalent to
\begin{eqnarray}
0 &=&\partial _{\rho }\left[ \rho \left( A^{j}\partial _{j}A^{\rho }-\rho
(A^{\varphi })^{2}\right) -\rho A^{\rho }\left( \partial _{\rho }A^{\rho
}+\partial _{z}A^{z}+\frac{1}{\rho }\,A^{\rho }\right) \right]  \notag \\
&&+\partial _{z}\left[ \rho A^{j}\partial _{j}A^{z}-\rho A^{z}\left(
\partial _{\rho }A^{\rho }+\partial _{z}A^{z}+\frac{1}{\rho }\,A^{\rho
}\right) \right]  \notag \\
&&+2M\rho \,\left( A^{\rho }\right) ^{2}+2M\rho ^{3}\,\left( A^{\varphi
}\right) ^{2}+2M\rho \,\left( A^{z}\right) ^{2}\,.
\end{eqnarray}%
Finally, replacing our solution \eqref{pre-sol}, keeping the constant $C$
for clarity, we find that all terms cancel out and again we obtain that the equation is satisfied only if the integration constant vanishes,
\begin{equation}
0=C^{2}\,.
\end{equation}

We conclude that the final solution is given by Eqs.~\eqref{sol}. This
solution is not satisfactory for two reasons. First, because $A^{i}$ is a complex
vector. Second, the function $Z(z)$ remains undetermined, showing that this branch does not give a unique solution for given boundary conditions. We
conclude that the holographic dislocation does not exist when the only torsion
component is the horizontal axial torsion field.

\subsubsection{Non-horizontal axial torsion \texorpdfstring{$A_{z}\neq 0$}{Az=0}}
\label{Nonhorizontal}

Now we consider the case of the vertical axial torsion, that is, $A_{z}\neq 0
$. We need to solve linear differential equations
\begin{eqnarray}
0 &=&\bar{C}_{1}\propto A^{i}\partial _{i}A^{\rho }-\rho (A^{\varphi
})^{2}\,,  \notag \\
0 &=&\bar{C}_{2}\propto A^{i}\partial _{i}\left( \rho A^{\varphi }\right)
+A^{\varphi }A^{\rho }\,,  \notag \\
0 &=&\bar{C}\propto A^{i}\partial _{i}A^{z}\,,  \notag \\
0 &=&C_{1}\propto \left( \partial _{\rho }A^{\rho }+\partial _{z}A^{z}+\frac{%
1}{\rho }\,A^{\rho }\right) A^{\rho }+2A^{i}\partial _{i}A^{\rho }-2\rho
(A^{\varphi })^{2}\,, \\
0 &=&C_{2}\propto \left( \partial _{\rho }A^{\rho }+\partial _{z}A^{z}+\frac{%
1}{\rho }\,A^{\rho }\right) \rho A^{\varphi }+2A^{i}\partial _{i}\left( \rho
A^{\varphi }\right) +2A^{\varphi }A^{\rho }\,,  \notag \\
0 &=&C_{3}\propto \left( \partial _{\rho }A^{\rho }+\partial _{z}A^{z}+\frac{%
1}{\rho }\,A^{\rho }\right) A^{z}+2A^{i}\partial _{i}A^{z}\,,  \notag
\end{eqnarray}%
and also%
\begin{eqnarray}
0 &=&C\propto \partial _{\rho }\left[ \rho \left( A^{j}\partial _{j}A^{\rho
}-\rho (A^{\varphi })^{2}\right) -\rho A^{\rho }\left( \partial _{\rho
}A^{\rho }+\partial _{z}A^{z}+\frac{1}{\rho }\,A^{\rho }\right) \right]
\notag \\
&&+\partial _{z}\left[ \rho A^{j}\partial _{j}A^{z}-\rho A^{z}\left(
\partial _{\rho }A^{\rho }+\partial _{z}A^{z}+\frac{1}{\rho }\,A^{\rho
}\right) \right]   \notag \\
&&+2\rho \,\left( A^{\rho }\right) ^{2}+2\rho ^{3}\,\left( A^{\varphi
}\right) ^{2}+2\rho \,\left( A^{z}\right) ^{2}\,.
\end{eqnarray}%
Replacing $\bar{C}=0$ in $C_{3}=0$ and assuming $A^{z}\neq 0$, we obtain
\begin{equation}
\partial _{\rho }A^{\rho }+\partial _{z}A^{z}+\frac{1}{\rho }\,A^{\rho }=0\,,
\end{equation}%
in which case $C_{1}=0$ takes the form
\begin{equation}
A^{\rho }\partial _{\rho }A^{\rho }+A^{z}\partial _{z}A^{\rho }-\rho
(A^{\varphi })^{2}=0\,.
\end{equation}%
Plugging in all the known quantities in $C=0$, we find
\begin{equation}
0=\left( A^{\rho }\right) ^{2}+\rho \left( A^{\varphi }\right) ^{2}+\left(
A^{z}\right) ^{2}\,.
\end{equation}%
The only real solution of the above equation is $A^{i}=0$. Even allowing the complex values of $A^{i}$, the solution has two arbitrary functions, making it not determined, which is not physical.

We conclude that the absence of diagonal torsion leads only to non
physical solutions for the torsion field, or the ones without a dislocation.


\bibliography{bibCSWeyl}
\bibliographystyle{utphys}

\end{document}